\documentclass[12pt,prd,aps,amssymb,amsmath,tightenlines,showpacs]{article}
\usepackage[utf8]{inputenc}
\usepackage[a4paper,top=3cm,bottom=3cm,left=1.5cm,right=1.5cm]{geometry}
\usepackage{amssymb}
\usepackage{amsmath}
\usepackage{amsthm}
\usepackage{physics}
\usepackage{mathtools}
\usepackage{tipa}
\usepackage{latexsym} 
\usepackage{graphicx}
\usepackage{slashed}
\usepackage{bbold}
\usepackage{cancel}
\usepackage{comment}
\usepackage{color}
\usepackage{graphicx,epsfig,color}
\usepackage[dvipsnames]{xcolor}
\usepackage{comment}
\usepackage{cite}
\usepackage{tikz}
\usepackage[compat=1.1.0]{tikz-feynman}
\usepackage{caption}
\usepackage{subcaption}

\newcommand{\be}{\begin{equation}}
	\newcommand{\ee}{\end{equation}}
\newcommand{\bea}{\begin{eqnarray}}
	\newcommand{\eea}{\end{eqnarray}}
\newcommand{\vv}{``}

\newcommand{\mpl}{M_P}

\begin{document}
	\graphicspath{{FIGURE/}}
	\topmargin=-1cm
	
	\begin{center}
		\vskip 5pt
		{\Large{\bf Path integral measure and RG equations for gravity}}
		
		\vspace*{0.8 cm}
		
		Carlo Branchina\footnote{carlo.branchina@unical.it}\label{one}$^{\dagger}$,
		Vincenzo Branchina\footnote{branchina@ct.infn.it}\label{two}$^\ddagger$, 
		Filippo
		Contino\footnote{filippo.contino@ct.infn.it}\label{three}$^\ddagger$,
		Arcangelo
		Pernace\footnote{arcangelo.pernace@ct.infn.it}\label{four}$^\ddagger$
		\vspace*{0.4cm}

		{ \it ${}^\dagger$Department of Physics, University of Calabria, and INFN-Cosenza\\
			Arcavacata di Rende, I-87036, Cosenza, Italy 
			
			\vskip 5pt
			
			${}^\ddagger$Department of Physics, University of Catania, and INFN-Catania\\
			Via Santa Sofia 64, I-95123 
			Catania, Italy}
			
	 \vskip 20pt
	 {\bf Abstract}
		
\end{center}

{\small 

\noindent
Considering the Einstein-Hilbert truncation for the running action in (euclidean) quantum gravity, we derive the renormalization group equations for the cosmological and Newton constant. We find that these equations admit only the Gaussian fixed point with a UV-attractive and a UV-repulsive eigendirection, and that there is no sign of 
the non-trivial UV-attractive fixed point of the asymptotic safety scenario. Crucial to our analysis is a careful treatment of the measure in the path integral that defines the running action and a proper introduction of the physical running scale $k$. We also show why and how in usual implementations of the RG equations the aforementioned UV-attractive fixed point is generated.}

\section{Introduction}
\label{introductionRG}

Several attempts have been made towards the formulation of a quantum theory of gravity. On the field theoretic side, the typical approach is based on the use of path integrals in the Euclidean formulation, where the metric $g_{\mu\nu}$ is a quantum field as any other matter field. This theory, however, is not renormalizable by power counting, and this essentially suggests two possible scenarios \cite{Weinberg:1996kw}. Either it is valid up to a certain scale (say the Planck scale $\mpl$), above which it is replaced by a UV (ultraviolet) completion of different nature (possibly string theory), or it is non-perturbatively renormalizable through the existence of a UV-attractive fixed point with finite dimensional critical surface, a possibility that S. Weinberg dubbed asymptotic safety scenario \cite{Weinberg:1976xy}. 

The question of whether this scenario could be realized in quantum gravity was first examined by Weinberg himself. Resorting to dimensional regularization, he considered the theory in $d=2+\epsilon$ dimensions \cite{Weinberg:1976xy, Weinberg:1979} and noted that for small values of $G$ the beta function is $\beta(G,\epsilon)=\epsilon\,G-b\,G^2+\mathcal{O}(G^3)$ and that for $b>0$ (and sufficiently small values of $\epsilon$) a UV-attractive $\mathcal{O}(\epsilon)$ fixed point exists. Inspired by the work of Wilson and Fisher \cite{Wilson:1971dc} (who calculated the critical exponents for the scalar theory in $d=3$ dimensions by working in $4-\epsilon$ dimensions and expanding the results around $\epsilon=1$), he wondered on the possibility of getting results in $d=4$ dimensions expanding those obtained in $d=2+\epsilon$ around $\epsilon=2$ \cite{Weinberg:1976xy}. However, Weinberg showed that this program cannot be pursued\cite{Weinberg:1979} since in $d=4$ dimensions polar singularities appear that cannot be cancelled by counterterms contained in the lagrangian of the original $(2+\epsilon)$-dimensional theory. He then observed that the possibility of an asymptotic safety scenario should be investigated directly in $d=4$ dimensions \cite{Weinberg:1979}, although he considered the existence of new UV physics (string theory, and in any case not a QFT) as the most probable scenario \cite{Weinberg:1996kw}. 

Investigations on the possibility of realizing asymptotic safety directly in $d=4$ dimensions were started in the late nineties \cite{Reuter:1996cp, Souma:1999at, Reuter:2001ag} (see \cite{video} for a popular introduction to the asymptotic safety scenario) where the renormalization group (RG) approach to quantum gravity was implemented resorting to the effective average action formalism introduced in \cite{Wetterich:1989xg}. Considering the Einstein-Hilbert truncation, in addition to the Gaussian fixed point a non-trivial UV-attractive one $\lambda_{_\text{\tiny FP}}>0\,,\, g_{_\text{\tiny FP}}>0$  was found \cite{Reuter:2001ag}. Here $\lambda\equiv\Lambda_k/k^2$ and $g\equiv k^2\,G_k$ are the dimensionless running cosmological and Newton constant, with $\Lambda_k$ and $G_k$ the corresponding dimensionful parameters and $k$ the running scale. Successively, the existence of this non-trivial fixed point was confirmed resorting to the proper-time formalism \cite{Bonanno:2004sy}. 

In this work, we derive the RG equations for $\Lambda_k$ and $G_k$ paying attention to aspects that turn out to be crucial in the calculation of the one-loop effective action \cite{noi}, that were unfortunately missed in previous literature. More precisely: (i) we take into account all the terms in the measure that appears in the path integral that defines the running action (usually, when going from the Hamiltonian to the Lagrangian formalism, terms that come from the integration over conjugate momenta are either neglected or not fully taken into account); (ii) we properly identify the physical running scale $k$. This leads to flow equations for $\Lambda_k$ and $G_k$ that differ substantially from those of \cite{Reuter:1996cp, Reuter:2001ag} and/or \cite{Bonanno:2004sy}. We find that only the Gaussian fixed point is present and that there is no sign of the non-trivial UV-attractive fixed point of the asymptotic safety scenario. We also show that (and how) this latter fixed point is artificially generated in \cite{Reuter:1996cp,Souma:1999at,Reuter:2001ag} and \cite{Bonanno:2004sy}.

The rest of the paper is organized as follows. To pave the way to our analysis, in section\,\ref{setup} we briefly recall the main steps of the calculation of the one-loop effective action in the Einstein-Hilbert truncation \cite{noi}. In section\,\ref{derivationRGeq}, we derive and solve the RG equations for the running Newton and cosmological constant. Section\,\ref{fixedpoints} is devoted to the search for the fixed points. In section\,\ref{Comparison} we explain why in the usual realization of the RG flow the UV-attractive fixed point of the asymptotic safety scenario is artificially generated. Section\,\ref{conclusions} is for the conclusions.

\section{General setup. One-loop effective action}
\label{setup}

To setup the tools for our analysis, in the present section we briefly recall the main results of a recent paper \cite{noi}, where the one-loop effective action for pure gravity in the Einstein-Hilbert truncation was calculated paying due attention to the role played by the path integral measure.

Considering the (euclidean) gravitational action\footnote{The Einstein-Hilbert truncation is well justified since in the cosmological framework we only need to consider manifolds with typical length scale $l$ much larger than the Planck length, $l\gg\mpl^{-1}$, that in turn implies $\Lambda_{\rm cc}\ll\mpl^2$.}
\begin{equation}
	S[\,g_{\mu\nu}]=\frac{1}{16\pi G}\int\dd[4]x\,\sqrt{g}\,\left(-R+2\Lambda_{\rm cc}\right)\,,
	\label{bareaction*}
\end{equation}
in \cite{noi} the one-loop correction $\delta S^{1l}$ was calculated according to the geometrical construction of Vilkovisky \cite{Vilkovisky:1984st} and DeWitt \cite{DeWitt:1987te}, that allows to get a gauge invariant result for the effective action even off-shell. To this end, the background field method \cite{Abbott:1980hw,Abbott:1981ke} was used, and the metric $g_{\mu\nu}$ was written as the sum of a background $\Bar g_{\mu\nu}$ plus the fluctuation $h_{\mu\nu}$. More specifically, we considered a spherical background, $\Bar g_{\mu\nu}=g^{(a)}_{\mu\nu}$ ($a$ radius of the sphere),
\begin{equation}\label{decomp}
	g_{\mu\nu}=g^{(a)}_{\mu\nu}+h_{\mu\nu}\,.
\end{equation}
For $g_{\mu\nu}=g^{(a)}_{\mu\nu}$, the classical action\,\eqref{bareaction*} reads
\begin{equation}
	S^{(a)}=\frac{\pi\Lambda_{\rm cc}}{3G}a^4-\frac{2\pi}{G}a^2\,.
	\label{Sb}
\end{equation}
Following \cite{TaylorVeneziano}, the one-loop corrections to $\frac{\Lambda_{\rm cc}}{G}$ and $\frac{1}{G}$ are identified with the coefficients of the $a^4$ and $a^2$ terms in $\delta S^{1l}$, respectively.

As shown in \cite{FradkinTseytlin}, if the background metric has spherical symmetry, the one-loop Vilkovisky-DeWitt effective action coincides with the standard one calculated with the gauge-fixing term
\begin{equation}\label{gf}
	S_{\rm gf}=\frac{1}{32\pi G\xi}\int\dd[4]{x}\sqrt{g^{(a)}}\left[\nabla_\mu\left(h^\mu_{\nu}-\frac12\delta^\mu_{\nu}\,h^{\sigma}_{\sigma}\right)\right]\,,
\end{equation}
whose corresponding ghost action is
\begin{equation}\label{ghostaction}
	S_{\rm ghost}=\frac{1}{32\pi G}\int\dd[4]{x}\sqrt{g^{(a)}}\,g^{(a)\,\mu\nu}\,v_\mu^*\left(-\nabla_\rho\nabla^{\rho}-\frac{3}{a^2}\right)v_\nu\,,
\end{equation}
after taking the limit $\xi\to0$ at the end of the calculation. The one-loop correction $\delta S^{1l}$ is then obtained from
\begin{align}\label{effacgrav1}
	e^{-\delta S^{1l}}=\lim_{\xi\to0}\int\big[\mathcal{D}u(h)\mathcal{D}v_\rho^*\,\mathcal{D}v_\sigma\big]\, e^{-(S_2+S_{\rm gf}+S_{\rm ghost})}\,,
\end{align}
where $S_2$ is the quadratic term in the expansion of $S[g^{(a)}_{\mu\nu}+h_{\mu\nu}]$ around $g^{(a)}_{\mu\nu}$, and the measure $\big[\mathcal{D}u(h)\mathcal{D}v_\rho^*\,\mathcal{D}v_\sigma\big]$ is
\begin{equation}\label{meas}
	\big[\mathcal{D}u(h)\mathcal{D}v_\rho^*\,\mathcal{D}v_\sigma\big]\equiv \prod_x\Big[g^{(a)\,00}(x)\,\left(g^{(a)}(x)\right)^{-1}\Big(\prod_{\alpha\leq\,\beta}\dd{h_{\alpha\beta}(x)}\Big)\Big(\prod_{\rho}\dd{v_\rho^*(x)}\Big)\Big(\prod_{\sigma}\dd{v_\sigma(x)}\Big)\Big]\,.
\end{equation}
It is important to stress the presence of the terms $g^{(a)\,00}(x)\,\left(g^{(a)}(x)\right)^{-1}$ in the measure\,\eqref{meas} that, as shown by Fradkin and Vilkovisky in \cite{Fradkin:1973wke}, come from the integration over conjugate momenta in the original Hamiltonian formulation of the theory. 
The impact of these terms on the calculation of $\delta S^{1l}$ is conveniently taken into account introducing the dimensionless fields \,$\widehat h_{\mu\nu}(x)\equiv\left(32\pi G\right)^{-1/2}a^{-1}h_{\mu\nu}$, \,$\widehat v_\mu\equiv\left(32\pi G\right)^{-\frac12}v_\mu$ and $\widehat v_\mu^{\,*}\equiv\left(32\pi G\right)^{-\frac12}v_\mu^*$, and decomposing $\widehat h_{\mu\nu}(x)$ in the basis for symmetric tensors and $\widehat v_\mu$ and $\widehat v_\mu^{\,*}$ in the basis for vectors. Both bases are built in terms of the eigenfunctions of the {\it dimensionless} spin-$s$ Laplace-Beltrami operators $-\widetilde\square^{(s)}\equiv-a^2\square^{(s)}_{\,a}$ ($s=0,1,2$, and $-\square^{(s)}_{\,a}$ are the dimensionful operators for a sphere of radius $a$). The one-loop correction $\delta S^{1l}$ is then (inessential terms are omitted in the equation below. See \cite{noi} for details)
\begin{align}\label{result}
	\delta S^{1l}&=-\frac12\log\frac{\det_1 [-\widetilde\square^{(1)}-3]\det_2 [-\widetilde\square^{(0)}-6]}{\det_0 [-\widetilde\square^{(2)}-2a^2\Lambda_{\rm cc}+8]\det_2 [-\widetilde\square^{(0)}-2a^2\Lambda_{\rm cc}]}\,.
\end{align} 

In comparison with previous literature, the novelty of\,\eqref{result} is that, although the calculation is performed for a sphere of generic radius $a$, only the {\it dimensionless} Laplacians $-\widetilde\square^{(s)}$ appear. The reason why the fluctuation determinants in\,\eqref{result} turn out to be automatically dimensionless\footnote{There is no need to introduce any arbitrary scale $\mu$ to take care of dimensional arguments of the logarithms as usually done in the literature.} is that the Fradkin-Vilkovisky terms \cite{Fradkin:1973wke} in the path integral measure have been correctly taken into account (see\,\eqref{meas} and comments below). With few exceptions \cite{Fradkin:1976xa,Donoghue:2020hoh}, in previous literature these terms were mistreated\footnote{They were either missed, or different (incorrect) measure terms were used that do not come from the integration over conjugate momenta.}.
Had we missed these terms, the $a$-dependence of the determinants in\,\eqref{result} would have been altered. In particular, the argument of the logarithm would have been dimensionful, and this would have required the introduction of an arbitrary scale $\mu$ to make it dimensionless, which is what is typically done in the literature (see footnote 6).

The determinants ${\rm det}_i$ (Eq.\,\eqref{result}) are calculated following two different methods \cite{noi}: 
\vskip 4pt
\noindent
(i) {\it direct product of the eigenvalues} of the fluctuation operators;

\vskip 4pt
\noindent
(ii) {\it proper-time}. 

\vskip 4pt
Concerning the method (i), it is worth to recall that the dimensionless eigenvalues $\lambda_n^{(s)}$ of {\footnotesize $-\widetilde\square^{(s)}$} are ($D_n^{(s)}$ are the corresponding degeneracies)
\begin{equation}
	\lambda_n^{(s)}=n^2+3n-s\quad ,\quad D_n^{(s)}=\frac{2s+1}{3}\left(n+\frac{3}{2}\right)^3-\frac{(2s+1)^3}{12}\left(n+\frac{3}{2}\right)\,,\,\,\,\text{with}\,\,\,n=s,s+1,\dots
	\label{eigenvalues}
\end{equation}
Each determinant $\text{\rm det}_i$ in\,\eqref{result} is calculated via the direct product of eigenvalues, and made finite through the introduction of a numerical cut $N$ ($\gg1$) on their number (see Eq.\,\eqref{MpN} below for the connection between $N$ and the physical UV cutoff $\Lambda_{\rm cut}$, say for instance the Planck scale $\mpl$). The expression of $\delta S^{1l}$ in terms of $\lambda_n^{(s)}$ and $D_n^{(s)}$ is (below the subscript $N$ in $\delta S^{1l}_{\text{\tiny$N$}}$ is introduced to indicate that $\delta S^{1l}$ is calculated with the numerical cut $N$)
\begin{align}
	\delta S^{1l}_{\text{\tiny$N$}}=\frac12\sum_{n=2}^{N-2} &\Bigl[D_n^{(2)}\log\left(\lambda_n^{(2)}-2a^2\Lambda_{\rm cc}+8\right)+D_n^{(0)}\log\left(\lambda_n^{(0)}-2a^2\Lambda_{\rm cc}\right)\nonumber\\
	&-D_n^{(1)}\log\left(\lambda_n^{(1)}-3\right)-D_n^{(0)}\log\left(\lambda_n^{(0)}-6\right)\Bigr]\,.
	\label{calculation1}
\end{align}
Expanding for $N\gg1$, and omitting as before inessential terms, we have
\begin{align}
    \delta S^{1l}_{\text{\tiny$N$}}=\,&-\left(\Lambda_{\rm cc}^2 \log N^2\right)a^4+ \Lambda_{\rm cc}\left(-N^2+8\log N^2\right)a^2\nonumber\\
	&+\frac{N^4}{24}\left(-1+2\log N^2\right)+\frac{N^2}{36}\left(203-75\log N^2\right)-\frac{779}{90}\log N^2\,.
	\label{1lres}
\end{align}
From\,\eqref{1lres}, the cosmological and Newton constant at one-loop turn out to be
\begin{align}
	\Lambda^{1l}_{\rm cc}=&
	\frac{\Lambda_{\rm cc}\left(1-\frac{3G\Lambda_{\rm cc}}{\pi} \log N^2\right)}{1+\frac{G\Lambda_{\rm cc}}{2\pi}\left(N^2-8\log N^2\right)}\label{CC}\\
	G^{1l}=&
	\frac{G}{1+\frac{G\Lambda_{\rm cc}}{2\pi}\left(N^2-8\log N^2\right)}\,.
	\label{NC}
\end{align}
On physical grounds, we limit ourselves to consider only the cases in which $\Lambda_{\rm cc}^{1l}$ and $G^{1l}$ have the same sign (both positive) of the measured cosmological and Newton constant. A simple inspection of\,\,\eqref{CC} and\,\eqref{NC} shows that under this physical request only positive values of the bare parameters $\Lambda_{\rm cc}$ and $G$ are admitted. Sticking then to the case $\Lambda_{\rm cc}>0$, we observe that the classical (de Sitter) solution {\small $a_{_{\rm dS}}=\sqrt{3/\Lambda_{\rm cc}}$} (see Eq.\,\eqref{Sb}) is the (tree-level approximation to the) size of the universe. Therefore, the connection between $N$ and the physical cutoff scale $\Lambda_{\rm cut}$ (say $\mpl$) is (see \cite{noi} for further comments on this point)
\begin{equation}\label{MpN}
	\Lambda_{\rm cut}=\frac{N}{a_{_{\rm dS}}}=N\sqrt{\frac{\Lambda_{\rm cc}}{3}}\,.
\end{equation}

Moving now to the method (ii), the determinants ${\rm det}_i$ are calculated using a dimensionless proper-time $\tau$ with lower integration bound {\small $1/N^2$} (below $\alpha=3,\,6,\,2a^2\Lambda_{\rm cc}-8,\,2a^2\Lambda_{\rm cc}$, see\,\eqref{result}) 
\begin{equation}\label{propertime*}
	\qquad\qquad\qquad\quad{\rm det}_{i}(-\widetilde\square^{(s)}-\alpha)=e^{-\int_{1/N^2}^{+\infty}\frac{\dd{\tau}}{\tau}\,{\rm K}_i^{(s)}(\tau)}\, \quad\quad \text{\footnotesize $\Big({\rm K}_i^{(s)}(\tau)= \sum_{n=s+i}^{+\infty}D^{(s)}_n\, e^{-\tau\left(\lambda^{(s)}_n-\alpha\right)}\Big)$}\,.
\end{equation}
Expanding as before $\delta S^{1l}_{\text{\tiny$N$}}$ for $N\gg1$ (inessential terms again omitted)
\begin{align}
	\delta S^{1l}_{\text{\tiny$N$}}=\, &-\left(\Lambda_{\rm cc} ^2\log N^2\right)a^4+\Lambda_{\rm cc}\left(-N^2+8\log N^2\right)a^2\nonumber\\
	&-\frac{N^4}{12}+\frac{17 }{3}N^2-\frac{1859}{90}\log N^2\,.
	\label{oneloopresHK*}
\end{align}
The relation between the numerical cut $N$ and the physical cutoff $\Lambda_{\rm cut}$ is as in\,\eqref{MpN}.

It is immediately seen from\,\eqref{1lres} and\,\eqref{oneloopresHK*} that both methods give the same one-loop result for the vacuum energy $\rho_{\rm vac}=\frac{\Lambda_{\rm cc}}{8\pi G}$ and for the inverse Newton constant $\frac{1}{G}$. In particular, the one-loop correction to $\rho_{\rm vac}$ contains only a (mild) logarithmic sensitivity to the UV scale. This result is at odds with the usual one (which is obtained resorting to proper-time regularization within the heat kernel expansion) where quartic and quadratic divergences are also present. As shown in \cite{noi}, these latter divergences are spurious since they come from an incorrect identification of the physical cutoff $\Lambda_{\rm cut}$ with $N/a$ rather than with $N/a_{_{\rm dS}}$ (see\,\eqref{MpN}). In fact, since the physical cutoff $\Lambda_{\rm cut}$ is obviously fixed, the identification with $N/a$ would imply that the number $N$ of eigenvalues retained in the calculation of the fluctuation determinants in\,\eqref{result} would change with varying $a$. This would alter the coefficients of $a^4$ and $a^2$ in $\delta S^{1l}_{\text{\tiny $N$}}$, thus artificially modifying the one-loop corrections to the vacuum energy and Newton constant. In particular, this incorrect identification of $\Lambda_{\rm cut}$ is at the origin of the typically acknowledged strong power-like UV sensitivity of the vacuum energy.

With the help of the results and techniques presented in this section, we now proceed to the derivation of the RG equations for the cosmological and Newton constant. 

\section{RG equations for the cosmological and Newton constant}
\label{derivationRGeq}

In this section we derive the RG equation for pure gravity, considering the Einstein-Hilbert truncation\,\eqref{bareaction*}. 
As for the one-loop calculation presented in the previous section, we write the metric $g_{\mu\nu}$ as the sum of a spherical background $g^{(a)}_{\mu\nu}$ plus the fluctuation $h_{\mu\nu}$ (see\,\eqref{decomp}). 
Our RG strategy is as follows. The (bare) action $S_{\text{\tiny $N$}}$ at the \vv UV scale" $N$, that in terms of the dimensionful physical cutoff $\Lambda_{\rm cut}$ ($=N/a_{\text{\tiny dS}}$, see\,\eqref{MpN}) can be indicated as $S_{\text{\tiny $\Lambda_{\rm cut}$}}$, contains modes of\, $\widehat h_{\mu\nu}$ up to the $N$-th ones. 
The action $S_{\text{\tiny$L$}}$ at the \vv lower scale" $L$ ($<N$) is obtained integrating out the modes within the range $[L,N]$. The action $S_{\text{\tiny$L$$-$$\delta L$}}$ at the scale $L-\delta L$ is obtained from $S_{\text{\tiny$L$}}$ integrating out the modes in the \vv infinitesimal shell" $[L-\delta L, L]$ (\text{\small $\frac{\delta L}{L}\ll1$})
\begin{equation}
	S_{\text{\tiny$L$$-$$\delta L$}}[\,g_{\mu\nu}^{(a)}]=S_{\text{\tiny$L$}}[\,g_{\mu\nu}^{(a)}]+\delta S_{\text{\tiny$L$}}\,,
	\label{RG1generic}
\end{equation}
where $\delta S_{\text{\tiny$L$}}$ is given by Eq.\,\eqref{result} with $\Lambda_{\rm cc}$ replaced by $\Lambda_{\text{\tiny$L$}}$ and the fluctuation determinants restricted to the subspace of modes in the shell\footnote{For each shell $[L-\delta L,L]$, the contribution of the gauge-fixing and ghost terms is taken into account considering in $S_{\rm gf}$ and $S_{\rm ghost}$ only the modes within such a shell.} $[L-\delta L,L]$.

Following steps similar to those presented in the previous section, we now derive $\delta S_{\text{\tiny$L$}}$ resorting to two different methods: (i) direct sum over the eigenvalues of the fluctuation operators; (ii) proper-time. For $S_{\text{\tiny$L$}}$ we take the Einstein-Hilbert truncation ansatz 
\begin{equation}
	S_{\text{\tiny$L$}}=\frac{1}{16\pi G_{\text{\tiny$L$}}}\int\dd[4]x\,\sqrt{g}\,\left(-R+2\Lambda_{\text{\tiny$L$}}\right)\,\,\underset{\text{$\,g_{\mu\nu}=g^{(a)}_{\mu\nu}$}}{=}\,\,\,\frac{\pi\Lambda_{\text{\tiny$L$}}}{3G_{\text{\tiny$L$}}}a^4-\frac{2\pi}{G_{\text{\tiny$L$}}}a^2\,,
	\label{runningaction}
\end{equation}
whose minimum is $a^{\text{\tiny dS}}_{\text{\tiny$L$}}=\sqrt{3/\Lambda_{\text{\tiny$L$}}}$. Note that, according to the notation introduced in the above equation, the bare action\,\eqref{bareaction*} is obtained for $L=N$ with
\begin{equation}\label{bareparameters}
	\Lambda_{\text{\tiny$N$}}\equiv\Lambda_{\rm cc}\qquad\text{and}\qquad G_{\text{\tiny$N$}}\equiv G\,.
\end{equation}
From the equation\,\eqref{RG1generic} for $S_{\text{\tiny$L$}}$ we will finally derive the RG equations for the running cosmological and Newton constant, $\Lambda_{\text{\tiny$L$}}$ and $G_{\text{\tiny$L$}}$ respectively. As we will see, both methods (i) and (ii) give rise to the same equations.

\subsection{Method (i): product of eigenvalues}
\label{RG-1}

In this case, $\delta S_{\text{\tiny$L$}}$ is obtained taking the product of eigenvalues of the fluctuation operators in the shell $[L-\delta L,L]$. We have
\begin{equation}
	S_{\text{\tiny$L$$-$$\delta L$}}[\,g_{\mu\nu}^{(a)}]=S_{\text{\tiny$L$}}[\,g_{\mu\nu}^{(a)}]+\delta S_{\text{\tiny$L$}}= S_{\text{\tiny$L$}}[\,g_{\mu\nu}^{(a)}]+\sum_{n\in\,[L-\delta L,L]}f_{_{\text{\tiny$L$}}}(n),
	\label{RG1}
\end{equation}
with $f_{_{\text{\tiny$L$}}}(n)$ given by (see Eq.\,\eqref{result} with $\Lambda_{\rm cc}$ replaced by $\Lambda_{\text{\tiny$L$}}$ and Eq.\,\eqref{eigenvalues})
\begin{align}\label{function}
	f_{_{\text{\tiny$L$}}}(n)&=D_n^{(2)}\log\left(\lambda_n^{(2)}-2a^2\Lambda_{\text{\tiny$L$}}+8\right)+D_n^{(0)}\log\left(\lambda_n^{(0)}-2a^2\Lambda_{\text{\tiny$L$}}\right)\nonumber\\
	&-\,D_n^{(1)}\log\left(\lambda_n^{(1)}-3\right)-D_n^{(0)}\log\left(\lambda_n^{(0)}-6\right)\,.
\end{align}
In differential form, Eq.\,\eqref{RG1} is written as\,\footnote{Note that the right hand side of\,\eqref{RG2} is nothing but the derivative with respect to $L$ of the one-loop contribution $\delta S^{1l}_{\text{\tiny$L$}}$ calculated with numerical cut $L$ (see\,\eqref{1lres}).}
\begin{align}
	\pdv{S_{\text{\tiny$L$}}}{L\,}=-\left(\pdv{}{L}\sum_{n=2}^{L-2}f_{_{\text{\tiny$L$}}}(n)\right)_{\Lambda_{\text{\tiny$L$}},\,G_{\text{\tiny$L$}}}\,,
	\label{RG2}
\end{align}
where the subscripts $\Lambda_{\text{\tiny$L$}}$ and $G_{\text{\tiny$L$}}$ indicate that the derivative with respect to $L$ is performed keeping $\Lambda_{\text{\tiny$L$}}$ and $G_{\text{\tiny$L$}}$ fixed. Note that the minimal value for $L$ is $L_{\rm min}=4$.
Eq.\,\eqref{RG2} describes the evolution of $S_{\text{\tiny$L$}}$ with the running $L$. 
The right hand side of\,\,\eqref{RG2} can be evaluated using the identity \,$\log\left(x/y\right)=-\int_{0}^{+\infty}\dd{z}\left[\left(x+z\right)^{-1}-\left(y+z\right)^{-1}\right]$. Expanding the result for\,\footnote{Since $L_{\rm min}=4$, $L \gg 1$ is realised practically in the whole range of $L$.} $L\gg1$, we get
\begin{align}
	L\,\pdv{S_{\text{\tiny$L$}}}{L\,}&=2\Lambda_{\text{\tiny$L$}}^2\,a^4+2\Lambda_{\text{\tiny$L$}}\left(L^2-8\right)a^2-\frac{L^2 (2L^2-25)}{6}  \log L^2-\frac{64L^2}{9} +\frac{779}{45}+\mathcal{O}\left(\frac{1}{L^2}\right)
	\,.
	\label{RG3}
\end{align}
Inserting in\,\eqref{RG3} the Einstein-Hilbert truncation\,\eqref{runningaction} for $S_{\text{\tiny$L$}}$, from the identification of the coefficients of $a^4$ and $a^2$ of the first member with those of the second member we get the RG equations 
\begin{align}
	L\,\dv{}{L}\frac{\Lambda_{\text{\tiny$L$}}}{G_{\text{\tiny$L$}}}=&\,\frac{6}{\pi}\Lambda_{\text{\tiny$L$}}^2\label{Lambda}\\
	L\,\dv{}{L}\frac{1}{G_{\text{\tiny$L$}}}=&-\frac{\Lambda_{\text{\tiny$L$}}}{\pi}\left(L^2-8\right)\,,
	\label{G}
\end{align}
that are easily translated into the equivalent equations
\begin{align}
	L\,\dv{\Lambda_{\text{\tiny$L$}}}{L}=&\,\frac{G_{\text{\tiny$L$}}\Lambda_{\text{\tiny$L$}}^2}{\pi}\left(L^2-2\right)\label{RGequations3_1}\\
	L\,\dv{G_{\text{\tiny$L$}}}{L}=&\,\frac{G_{\text{\tiny$L$}}^2\Lambda_{\text{\tiny$L$}}}{\pi}\left(L^2-8\right)\label{RGequations3_2}\,.
\end{align}
Before proceeding with the study of these equations, and search for their solution, it is convenient (the reason will be clear in the next subsection) to move first to the derivation of the RG equations resorting to the proper-time method. 

\subsection{Method (ii): proper-time}
\label{RG-2}

Let us move now to the proper-time method. In this case, $\delta S_{\text{\tiny$L$}}$ is given by\,\eqref{result} with $\Lambda_{\rm cc}$ replaced by $\Lambda_{\text{\tiny$L$}}$ and the derminants ${\rm det}_i$ calculated using\,\eqref{propertime*} with integration over the dimensionless proper-time $\tau$ restricted to the interval {\small $\left[1/L^2\,,\,1/(L-\delta L)^2\right]$}. Therefore, $\delta S_{\text{\tiny$L$}}$ is the combination of terms of the kind ($\alpha=3,\,6,\,2a^2\Lambda_{\text{\tiny$L$}}-8,\,2a^2\Lambda_{\text{\tiny$L$}}$, see\,\eqref{propertime*})
\begin{equation}
	\frac12\log{\rm det}_{i}(-\widetilde\square^{(s)}-\alpha)=-\frac12\int_{1/L^2}^{1/(L-\delta L)^2}\frac{\dd{\tau}}{\tau}\,{\rm K}_i^{(s)}(\tau)\, \quad\quad \text{\footnotesize $\Big({\rm K}_i^{(s)}(\tau)= \sum_{n=s+i}^{+\infty}D^{(s)}_n\, e^{-\tau\left(\lambda^{(s)}_n-\alpha\right)}\Big)$}\,.
\end{equation}
In differential form, the RG equation for the running action $S_{\text{\tiny$L$}}$ is  
\begin{align}
	\pdv{S_{\text{\tiny$L$}}}{L\,}=-\left(\pdv{\left(\delta S^{1l}_{\text{\tiny$L$}}\right)}{L}\right)_{\Lambda_{\text{\tiny$L$}},\,G_{\text{\tiny$L$}}}\,\,
	\label{RG2pt}
\end{align}
where $\delta S^{1l}_{\text{\tiny$L$}}$ is given by\,\eqref{oneloopresHK*} with $N$ replaced by $L$ and $\Lambda_{\rm cc}$ by $\Lambda_{\text{\tiny$L$}}$.
Performing the derivative in the right hand side of\,\eqref{RG2pt} ($\Lambda_{\text{\tiny$L$}}$ and $G_{\text{\tiny$L$}}$ fixed as in\,\eqref{RG2}), we finally get
\begin{align}
	L\,\pdv{S_{\text{\tiny$L$}}}{L\,}&=2\Lambda_{\text{\tiny$L$}}^2\,a^4+2\Lambda_{\text{\tiny$L$}}\left(L^2-8\right)a^2+\frac{L^4}{3}-\frac{34L^2}{3}+\frac{1859}{45}+\mathcal{O}\left(\frac{1}{L^2}\right)
	\,.
	\label{RG3*}
\end{align}

\noindent
Inserting the Einstein-Hilbert truncation\,\eqref{runningaction} in the left hand side of\,\eqref{RG3*}, and identifying the coefficients of $a^4$ and $a^2$ of the first and second member, we obtain the RG equations for $\Lambda_{\text{\tiny$L$}}/G_{\text{\tiny$L$}}$ and $1/G_{\text{\tiny$L$}}$. Remarkably, they turn out to be the same as those obtained resorting to the direct product of eigenvalues (previous section). Therefore, independently of the method used for their derivation, the RG equations for $\Lambda_{\text{\tiny$L$}}$ and $G_{\text{\tiny$L$}}$ turn out to be those in Eqs.\,\eqref{RGequations3_1} and\,\eqref{RGequations3_2}. 

Before ending this section, it is worth to stress that the proper-time formalism was already used to derive RG equations for the cosmological and Newton constant \cite{Bonanno:2004sy}, but, as we will see in section\,\ref{Ltok}, the RG equations of \cite{Bonanno:2004sy} are substantially different from ours. We will comment on this difference in this latter section and more in detail in section\,\ref{Comparison}. 

In the next section, we look for the solution to Eqs.\,\eqref{RGequations3_1} and\,\eqref{RGequations3_2}, and we will see that under a well controlled approximation they can be solved analytically. We will also solve them numerically, and compare the analytic and numerical results.

\subsection{Solution of the RG equations}
\label{solutions}

Let us consider the RG equations\,\eqref{RGequations3_1} and\,\eqref{RGequations3_2} for $\Lambda_{\text{\tiny$L$}}$ and $G_{\text{\tiny$L$}}$. For $L\gg1$ (remember that $L_{\rm min}=4$, so that this condition is verified for all but only few modes near $L_{\rm min}$) they can be approximated as
\begin{align}
	L\,\dv{\Lambda_{\text{\tiny$L$}}}{L}=&\,\frac{G_{\text{\tiny$L$}}\Lambda_{\text{\tiny$L$}}^2}{\pi}L^2\label{Lambdaapprox}\\
	L\,\dv{G_{\text{\tiny$L$}}}{L}=&\,\frac{G_{\text{\tiny$L$}}^2\Lambda_{\text{\tiny$L$}}}{\pi}L^2\label{Gapprox}\,.
\end{align}
The interest of this approximation is that Eqs.\,\eqref{Lambdaapprox} and\,\eqref{Gapprox} can be solved analytically. For the UV boundary conditions $\Lambda_{\text{\tiny $N$}}=\Lambda_{\rm cc}$ and $G_{\text{\tiny $N$}}=G$ the solution is
\begin{align}
	\Lambda_{\text{\tiny$L$}}=&\frac{\Lambda_{\rm cc}}{\sqrt{1+\frac{G\,\Lambda_{\rm cc}}{\pi}(N^2-L^2)}}\label{anLambda}\\
	G_{\text{\tiny$L$}}=&\frac{G}{\sqrt{1+\frac{G\,\Lambda_{\rm cc}}{\pi}(N^2-L^2)}}\,.
	\label{anG}
\end{align}
According to\,\eqref{anLambda} and\,\eqref{anG}, the sign of both $\Lambda_{\text{\tiny$L$}}$ and $G_{\text{\tiny$L$}}$ is fixed and given by the sign of $\Lambda_{\rm cc}$ and $G$ respectively. Moreover, from a simple inspection of\,\,\eqref{RGequations3_1} and\,\,\eqref{RGequations3_2} we see that this is not related to the approximation considered here, but holds true in general. On the contrary, while\,\eqref{anLambda} and \eqref{anG} predict that the vacuum energy $\rho_{_{\text{\tiny$L$}}}=\frac{\Lambda_{\text{\tiny$L$}}}{8\pi G_{\text{\tiny$L$}}}$ is constant (no running with $L$), from the numerical solution of\,\eqref{RGequations3_1} and\,\eqref{RGequations3_2} we see that instead this is due to the approximation. Nevertheless, the running of $\rho_{_{\text{\tiny$L$}}}$ is so slow that it is very well approximated by the constant behaviour given by\,\eqref{anLambda} and\,\eqref{anG}.

\begin{figure}[t]
	\centering
	\includegraphics[width=0.48\linewidth]{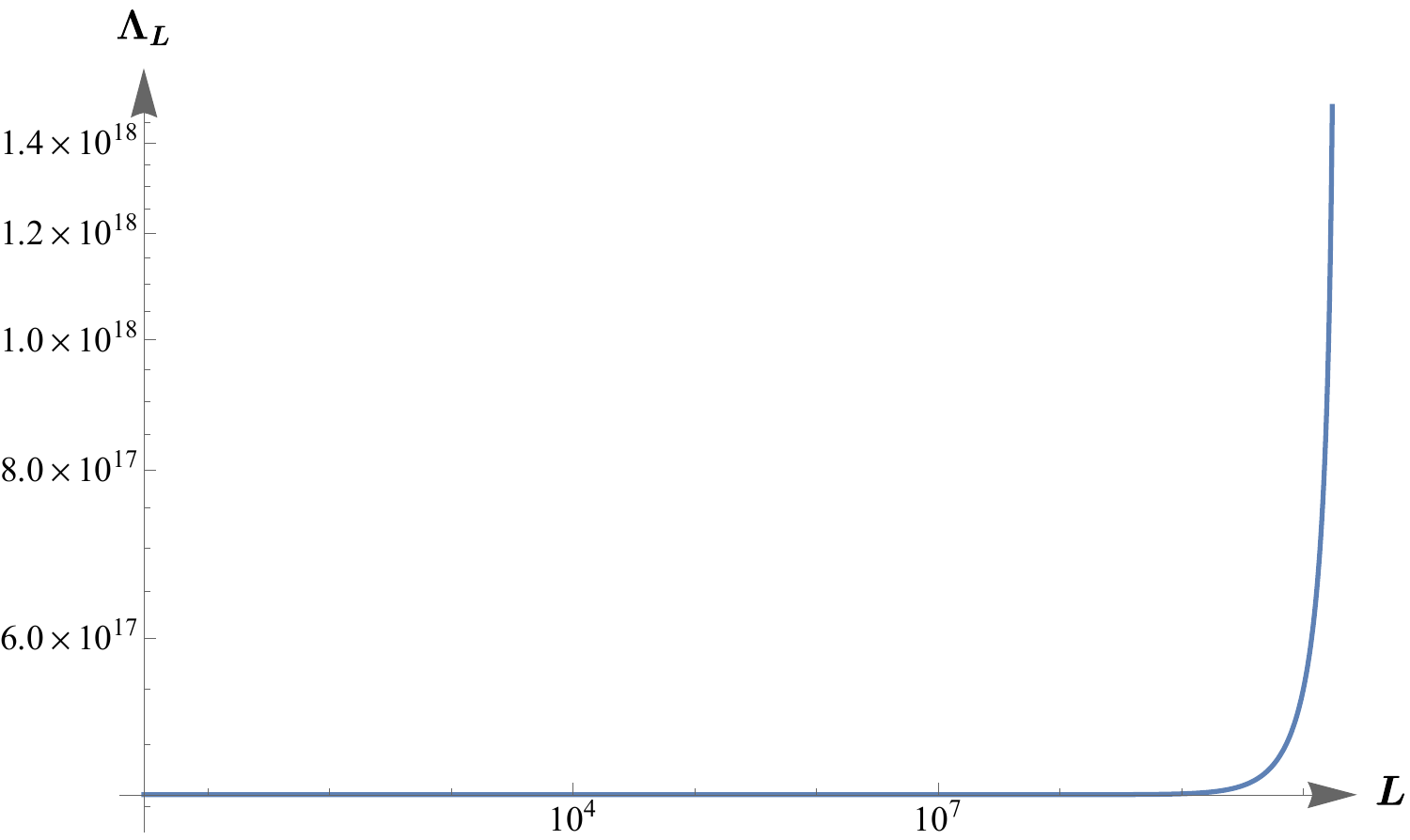}\hspace{15pt}
	\includegraphics[width=0.48\linewidth]{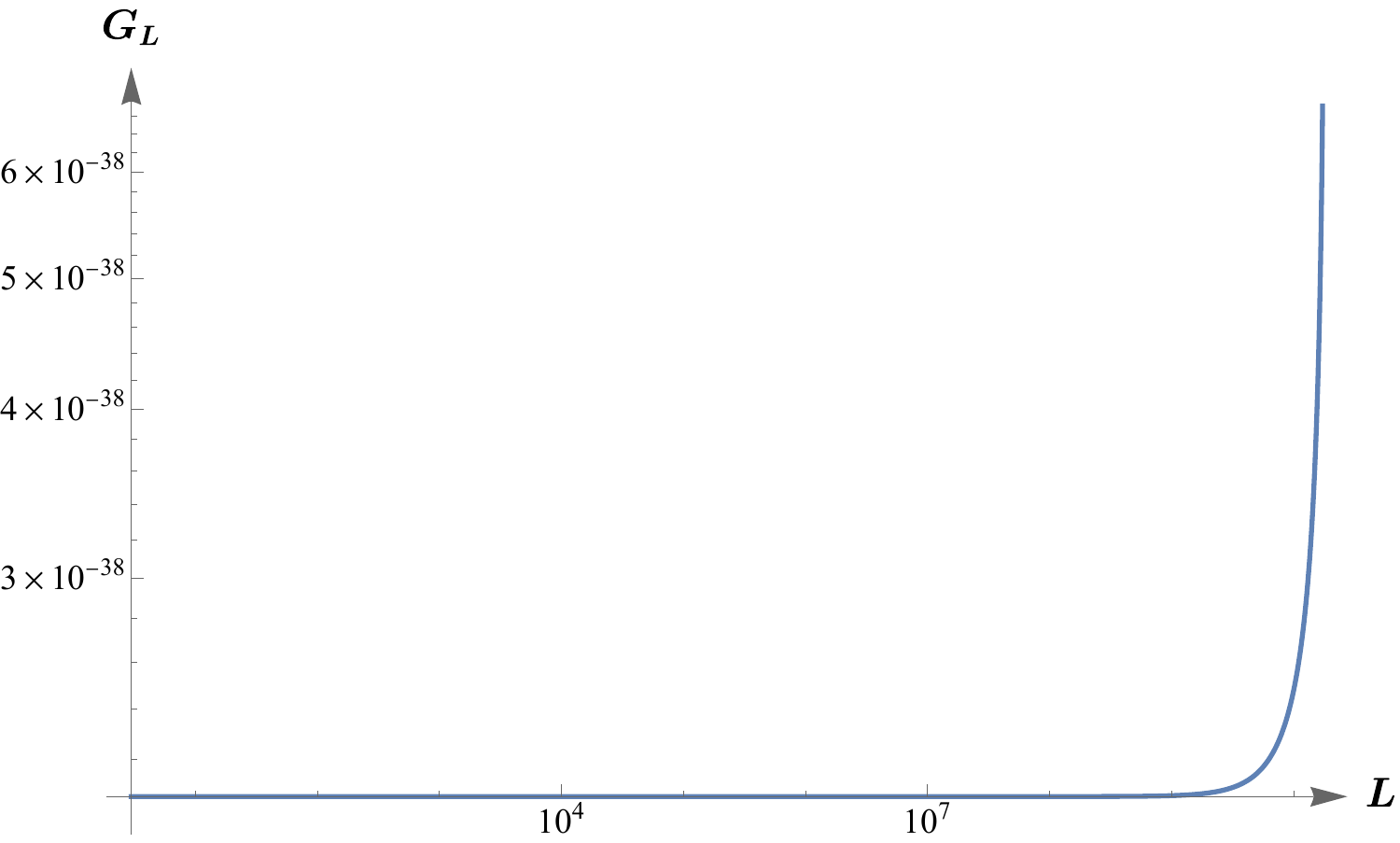}
	\caption[]{\footnotesize\textit{Left panel}: Log-log plot in the range $L_{\rm min}=4\leq L\leq 10^{10}$ of the cosmological constant flow\,\eqref{anLambda2}, for $\Lambda_{\rm cc}=10^{-20}\mpl^2$\,, $G=10\mpl^{-2}$ and $\Lambda_{\rm cut}=\mpl$. \textit{Right panel}: Log-log plot in the same range of $L$ of the Newton constant flow\,\eqref{anG2}, for the same values of $\Lambda_{\rm cut}$, $\Lambda_{\rm cc}$ and $G$. In both panels, $G$eV units are used.}
	\label{flowsL}
\end{figure}

In view of the above considerations, and since the measured values of the cosmological and Newton constant are both positive, from now on (unless explicitly stated) we restrict ourselves to consider positive UV boundaries $\Lambda_{\rm cc}>0$ and $G>0$. For completeness, in the Appendix we will also consider (and speculate on) the unphysical cases where one or both of them are negative. With the help of\,\,\eqref{MpN}, we now replace $N$ in\,\eqref{anLambda} and\,\eqref{anG} with the physical UV cutoff $\Lambda_{\rm cut}$, and get 
\begin{align}
	\Lambda_{\text{\tiny$L$}}=&\frac{\Lambda_{\rm cc}}{\sqrt{1+\frac{G}{\pi}(3\Lambda_{\rm cut}^2-\Lambda_{\rm cc} L^2)}}\label{anLambda2}\\
	G_{\text{\tiny$L$}}=&\frac{G}{\sqrt{1+\frac{G}{\pi}(3\Lambda_{\rm cut}^2-\Lambda_{\rm cc} L^2)}}\,.
	\label{anG2}
\end{align}
According to\,\eqref{anLambda2} and\,\eqref{anG2}, as long as $L$ is not much lower than $N$, the cosmological and Newton constant decrease for decreasing $L$ (the rapidity of the descent depends on the value of $G\Lambda_{\rm cc}$) and then practically freeze to the renormalized values 
\begin{align}
	\Lambda_{\text{\tiny IR}}&\sim \frac{\Lambda_{\rm cc}}{\sqrt{1+\frac{3G\Lambda_{\rm cut}^2}{\pi}}}\label{IRLambda}\\
	G_{\text{\tiny IR}}&\sim \frac{G}{\sqrt{1+\frac{3G\Lambda_{\rm cut}^2}{\pi}}}\label{IRG}\,.
\end{align}
This behaviour is seen in Fig.\,\ref{flowsL}, where we show a log-log plot of the flow\,\eqref{anLambda2} (left panel) and\,\eqref{anG2} (right panel) for $\Lambda_{\rm cut}=\mpl$, $\Lambda_{\rm cc}=10^{-20}\mpl^2$ and $G=10\,\mpl^{-2}$. 
Solving numerically equations\,\eqref{RGequations3_1} and\,\eqref{RGequations3_2} for different UV boundary values of $\Lambda_{\text{\tiny$L$}}$ and $G_{\text{\tiny$L$}}$, we see that\,\eqref{anLambda2} and\,\eqref{anG2} provide an excellent approximation to the exact solution. 

Eqs.\,\eqref{IRLambda} and\,\eqref{IRG} show an important outcome of our analysis. Since $\Lambda_{\rm cut}\sim\mpl$, for natural values of the Newton constant, i.e.\,\,$G\sim\mpl^{-2}$, from\,\eqref{IRLambda} and\,\eqref{IRG} we have that $\Lambda_{\text{\tiny IR}}\sim\Lambda_{\rm cc}$ and $G_{\text{\tiny IR}}\sim G$. This means that our RG  equations\,\eqref{RGequations3_1} and\,\eqref{RGequations3_2} give only a mild dressing of the cosmological and Newton constants. In other words, quantum fluctuations do not modify significantly the UV values $\Lambda_{\rm cc}$ and $G$: {\it no naturalness problem} arises in pure gravity\footnote{For recent discussions and different opinions on the cosmological constant naturalness problem in theories with compact extra dimensions see \cite{Montero:2022prj, Anchordoqui:2023laz, Branchina:2023rgi, Branchina:2023ogv, Branchina:2024ljd}. A novel approach to the naturalness problem for scalar theories is in \cite{Branchina:2022gll} (see also \cite{Branchina:2022jqc}).}. 

In the next section, we connect the \vv numerical scale" $L$ to the physical running scale $k$ and write the RG equations\,\eqref{RGequations3_1} and\,\eqref{RGequations3_2} in terms of $k$. In this new framework, we will partially repeat the study of the present section and add further comments. 

\subsection{From $L$ to the physical running scale $k$}
\label{Ltok}

According to\,\eqref{MpN}, the relation between the UV numerical cut $N$ and the physical cutoff $\Lambda_{\rm cut}$ is given by {\small $\Lambda_{\rm cut}=N/a_{\text{\tiny dS}}=N\sqrt{\Lambda_{\text{\tiny $N$}}/3}$}. Therefore, the relation between the numerical \vv running scale" $L$ and the physical running scale $k$ is
\begin{equation}\label{relationRG}
	k=\frac{L}{a^{\text{\tiny dS}}_{\text{\tiny$L$}}}
	\,,
\end{equation}
where $a^{\text{\tiny dS}}_{\text{\tiny$L$}}$ is the de Sitter radius that minimizes the action $S_{\text{\tiny$L$}}[g_{\mu\nu}^{(a)}]$ in\,\eqref{runningaction}
\begin{equation}\label{runningdS}
	a^{\text{\tiny dS}}_{\text{\tiny$L$}}=\sqrt{\frac{3}{\Lambda_{\text{\tiny$L$}}}}\,.
\end{equation}
As seen in the previous section, since we are considering positive UV boundaries\footnote{As already said, in the Appendix we will also consider (and speculate on) the unphysical cases where this condition is released.} $\Lambda_{\rm cc}>0$ and $G>0$, the running cosmological constant $\Lambda_{\text{\tiny$L$}}$ decreases monotonically for decreasing $L$ and remains positive in the whole range $L_{\rm min}\leq L\leq N$ ($L_{\rm min}=4$ is the minimal value of $L$, see\,\eqref{RG2}). Eq.\,\eqref{relationRG} establishes then a one-to-one correspondence between $L$ and $k$, with {\small $k\in[k_{_{\rm IR}},\Lambda_{\rm cut}]$} where
\begin{equation}\label{kIR}
	k_{_{\rm IR}}=\sqrt{\frac{16\Lambda_4}{3}\,}\,.
\end{equation}
With the help of\,\eqref{relationRG} and\,\eqref{runningdS}, Eqs.\eqref{RGequations3_1} and \eqref{RGequations3_2} can be written in terms of the running scale $k$ as ($\Lambda_k\equiv\Lambda_{\text{\tiny$L$}}$ and $G_k\equiv G_{\text{\tiny$L$}}$)
\begin{align}
	k\dv{\Lambda_k}{k}&=\frac{3 G_k}{\pi}\frac{\Lambda_k\left(k^2-\frac23\Lambda_k\right)}{1+\frac{3G_k}{2\pi}\left(k^2-\frac23\Lambda_k\right)}\label{Lambdaeq}\\
	k\dv{G_k}{k}&=\frac{3 G_k^2}{\pi}\frac{k^2-\frac83\Lambda_k}{1+\frac{3G_k}{2\pi}\left(k^2-\frac23\Lambda_k\right)}\label{Gequation}\,.
\end{align}

Before going on with our analysis, it is worth to remind what we found in the previous sections: the RG equations for the cosmological and Newton constant are\,\eqref{RGequations3_1} and\,\eqref{RGequations3_2} independently of the method used for their derivation (direct product of eigenvalues, section \ref{RG-1}; proper-time method, section \ref{RG-2}). Obviously, the same holds true for\,\eqref{Lambdaeq} and\,\eqref{Gequation}. Sticking to the proper-time method, we recall that this formalism was already used in \cite{Bonanno:2004sy} to derive RG equations for $\Lambda_k$ and $G_k$, but these equations are substantially different from our Eqs.\,\eqref{Lambdaeq} and\,\eqref{Gequation}. In section \ref{Comparison}, we will show what is at the origin of this difference, and explain why in our opinion\,\eqref{Lambdaeq} and\,\eqref{Gequation} are the correct RG equations for $\Lambda_k$ and $G_k$.

The system\,\eqref{Lambdaeq}-\eqref{Gequation} can be solved numerically, and the solutions are obviously the same as those found in the previous section for the system\,\eqref{RGequations3_1}-\eqref{RGequations3_2}, although given in terms of $k$, $\Lambda_k$ and $G_k$.
\begin{figure}[t]
	\centering
	\includegraphics[width=0.491\linewidth]{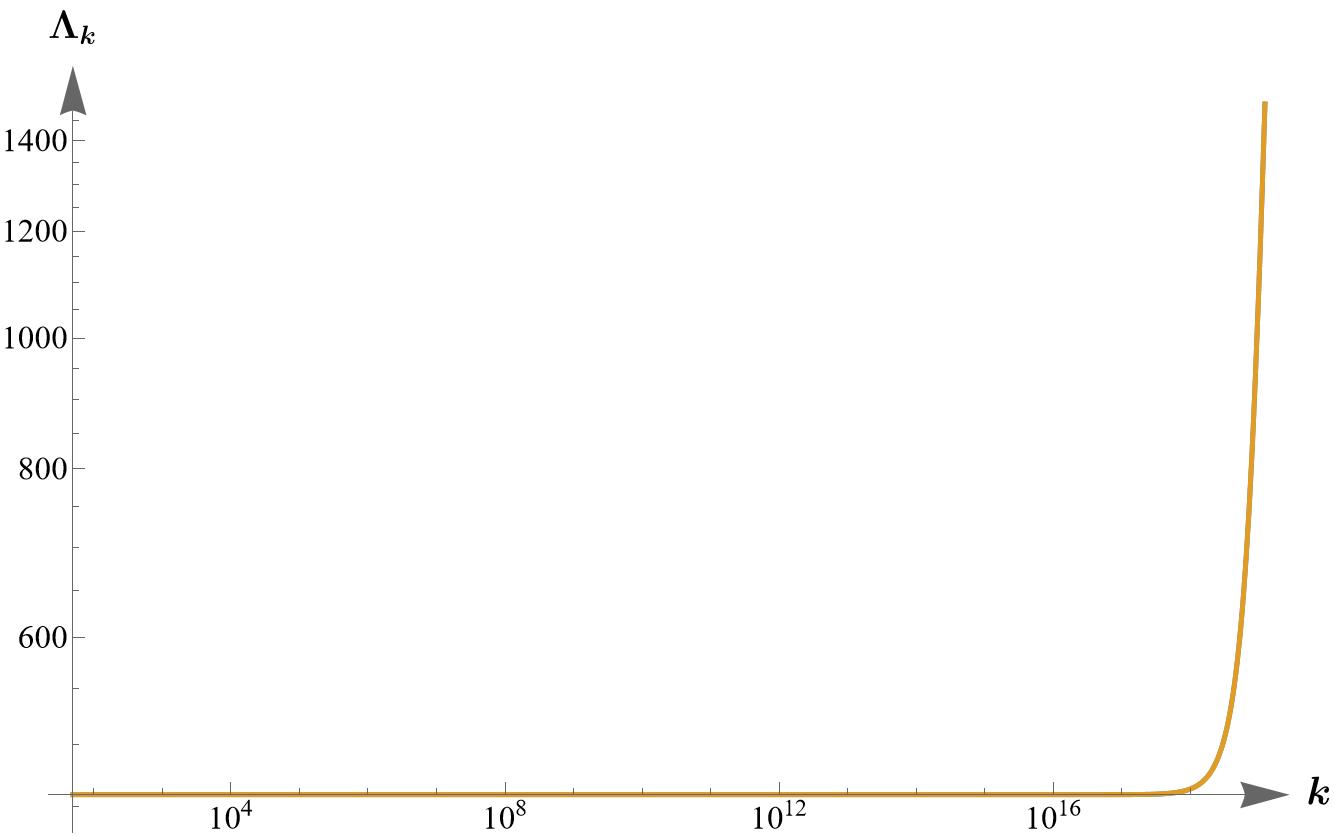}\hspace{5pt}
	\includegraphics[width=0.491\linewidth]{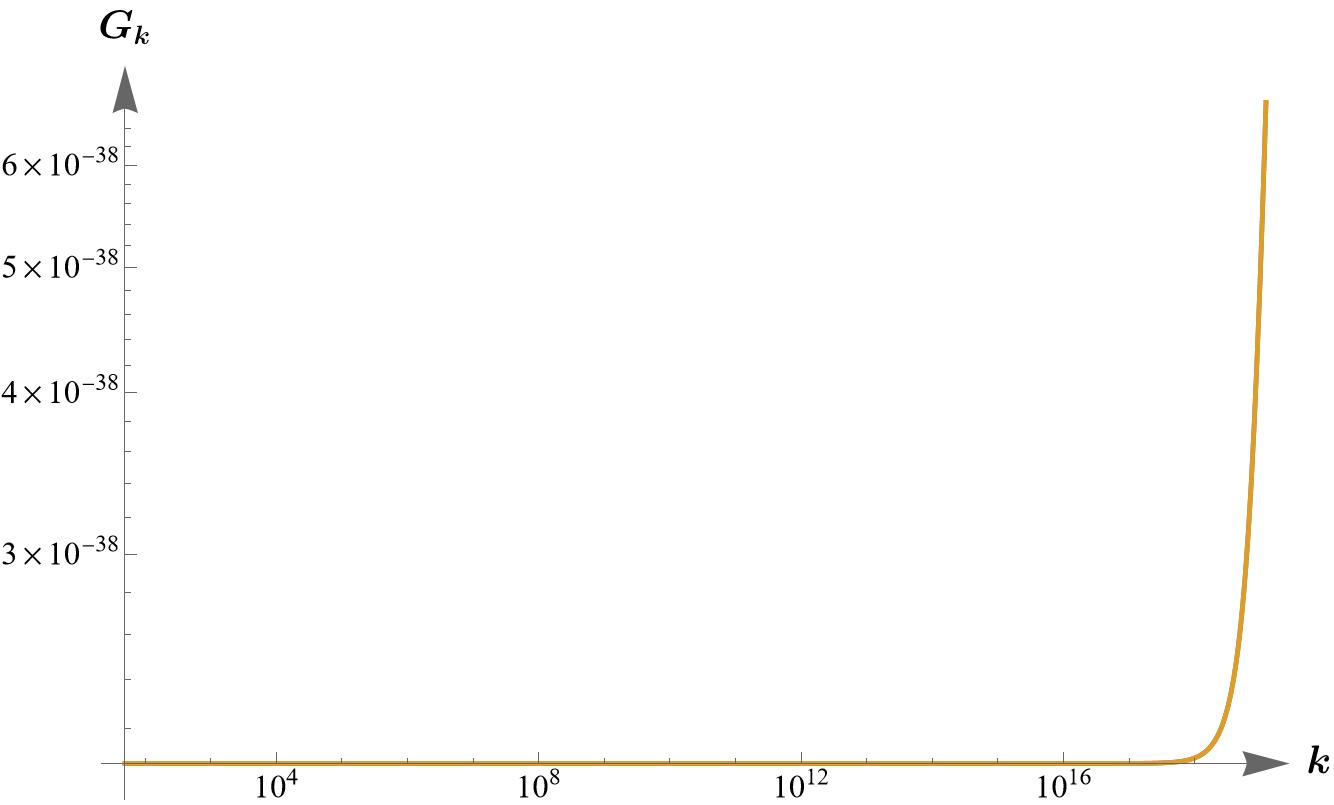}\hspace{15pt}
	\caption[]{\footnotesize \textit{Left panel}: Log-log plot of the cosmological constant flow ($G$eV units) in the range $k_{_{\rm IR}}\simeq 50\, G\text{eV}\leq k\leq \mpl$, for $\Lambda_{\rm cc}=10^{-35}\mpl^2$\,, $G=10\mpl^{-2}$ and $\Lambda_{\rm cut}=\mpl$. The approximated flow\,\eqref{solutionLambda} is plotted together with the corresponding (exact) flow from the numerical solution of\,\,\eqref{Lambdaeq}-\eqref{Gequation}. The two curves practically coincide. Around the scale $k\sim10^{18}\,G\text{eV}$ the flow creates an elbow that marks the transition from the $k^2$-running to a quick freezing to the value $\Lambda_{\text{\tiny IR}}\simeq459 (G\text{eV})^2$. \textit{Right panel}: Log-log plot of the Newton constant flow ($G$eV units) in the same range of $k$, and for the same $\Lambda_{\rm cut}$, $\Lambda_{\rm cc}$ and $G$. The approximated flow\,\eqref{solutionG} is plotted together with the corresponding (exact) flow from the numerical solution of\,\,\eqref{Lambdaeq}-\eqref{Gequation}. As for $\Lambda_k$, the two curves practically coincides, and around $k\sim10^{18}\,G\text{eV}$ the RG flow again forms an elbow that marks the transition from the $k^2$-running to the frozen value $G_{\text{\tiny IR}}\simeq2.07\cdot10^{-38} (G\text{eV})^{-2}$.}
	\label{AnandNumSol}
\end{figure}
Since  $\Lambda_k/k^2=3/L^2 \ll 1$ (see also footnote 5), we can expand\,\eqref{Lambdaeq} and\,\eqref{Gequation} around $\Lambda_k/k^2\sim0$. Retaining only the first non-trivial order in both equations we get
\begin{align}
	k\dv{\Lambda_k}{k}&=\frac{3 G_k}{\pi}\frac{k^2\Lambda_k}{1+\frac{3G_k}{2\pi}k^2}\label{Lambdaeq*}\\
	k\dv{G_k}{k}&=\frac{3 G_k^2}{\pi}\frac{k^2}{1+\frac{3G_k}{2\pi}k^2}\label{Gequation*}\,.
\end{align}
This system can be solved analytically. Taking at $k=\Lambda_{\rm cut}$ the UV boundary values $\Lambda_{\rm cc}$ and $G$ for $\Lambda_k$ and $G_k$ respectively, we get
\begin{align}
	\Lambda_k&=\frac{k^2\Lambda_{\rm cc}}{2\left(\Lambda_{\rm cut}^2+\frac{\pi}{3G}\right)}\left[1+\sqrt{1+\frac{4\pi}{3G}\left(\Lambda_{\rm cut}^2+\frac{\pi}{3G}\right)\frac{1}{k^4}}\,\right]\label{solutionLambda}\\
	G_k&=\frac{k^2G}{2\left(\Lambda_{\rm cut}^2+\frac{\pi}{3G}\right)}\left[1+\sqrt{1+\frac{4\pi}{3G}\left(\Lambda_{\rm cut}^2+\frac{\pi}{3G}\right)\frac{1}{k^4}}\,\right]\label{solutionG}\,.
\end{align}

In the left panel of Fig.\,\ref{AnandNumSol}, the analytic solution $\Lambda_k$ in\,\eqref{solutionLambda} is plotted (log-log plot) together with the corresponding numerical solution of the system\,\eqref{Lambdaeq}-\eqref{Gequation}, with $\Lambda_{\rm cut}=\mpl$, $\Lambda_{\rm cc}= 10^{-35}\mpl^2$ and $G= 10\mpl^{-2}$. The two curves are practically indistinguishable in the whole range $k_{_{\rm IR}}\leq k\leq \mpl$. A similar plot for $G_k$ is shown in the right panel. We verified that\,\eqref{solutionLambda} and\,\eqref{solutionG} very well approximate the numerical solutions to the full equations\,\eqref{Lambdaeq} and\,\eqref{Gequation} for different boundary values $\Lambda_{\rm cc}$ and $G$. From Fig.\,\ref{AnandNumSol} and from\,\eqref{solutionLambda} and\,\eqref{solutionG}, we see that $\Lambda_k$ and $G_k$ run quadratically with the scale $k$ until the latter reaches the \vv transition scale" 
\begin{equation}\label{elbow}
	k_{\rm tr}\sim \left(\frac{\Lambda_{\rm cut}^2+\frac{\pi}{3G}}{G}\right)^{1/4}\,.
\end{equation}
Around this scale, both $\Lambda_k$ and $G_k$ show an \vv elbow" that marks the transition from the quadratic running to a constant behaviour. This is seen in Fig.\,\ref{AnandNumSol}, where the elbow is in the region around the scale $k\sim10^{18}\,G\text{eV}$. Below $k_{\rm tr}$ the running cosmological and Newton constant rapidly freeze to their IR values \begin{align}\label{freezvalues}
\Lambda_{\text{\tiny IR}}=\frac{\Lambda_{\rm cc}}{\sqrt{1+\frac{3G\Lambda_{\rm cut}^2}{\pi}\,}\,} \qquad \qquad \qquad  G_{\text{\tiny IR}}=\frac{G}{\sqrt{1+\frac{3G\Lambda_{\rm cut}^2}{\pi}\,}\,}
\end{align} 
already found in the previous section (Eqs.\,\eqref{IRLambda} and\,\eqref{IRG}).
We observe that for $G\gtrsim \Lambda_{\rm cut}^{-2}\sim\mpl^{-2}$ the elbow appears at the very early stages of the UV running, i.e.\,\,$k_{\rm tr}\sim \mpl$ (see Fig.\,\ref{AnandNumSol}), which means that the running freezes very early. In other words, the quadratic evolution in $k$ is restricted to a very narrow window in the UV region.

In the next section we continue the study of the RG flow\,\eqref{Lambdaeq}-\eqref{Gequation}, and perform the fixed points analysis.

\section{Fixed points and RG flow}
\label{fixedpoints}

We introduce (as usual) the \vv RG time"  $t\equiv\log(k/k_0)$ ($k_0\leq\Lambda_{\rm cut}$ arbitrary scale) and the dimensionless running cosmological and Newton constant
\begin{equation}\label{dimlesscoup}
	\lambda(t)\equiv\Lambda_k/k^2 \quad\quad,\quad\quad g(t)\equiv k^2G_k\,.
\end{equation}
In terms of $\lambda$, $g$, the RG equations\,\eqref{Lambdaeq} and\,\eqref{Gequation} can be written as
\begin{align}
	\dv{\lambda}{t}&=-2\lambda+\frac{2 g\lambda\left(3-2\lambda\right)}{2\pi+g\left(3-2\lambda\right)}\equiv\beta_{\lambda}(\lambda,g)\label{dimlessLambdaeq}\\
	\dv{g}{t}&=\,\,\,\,\,2g+\frac{2g^2\left(3-8\lambda\right)}{2\pi+g\left(3-2\lambda\right)}\equiv\beta_{g}(\lambda,g)\label{dimlessGequation}\,.
\end{align}
The fixed points $(\lambda_i,g_i)$ are the solutions of \,$\beta_{\lambda}(\lambda,g)=0$ and $
\beta_{g}(\lambda,g)=0$. We find
\begin{align}\label{GFP}
\left(\lambda_1,g_1\right)&=\left(0,0\right)\\
\left(\lambda_2,g_2\right)&=\left(0,-\frac{\pi}{3}\right)\,.\label{NTFP}
\end{align}
In section\,\ref{solutions}, we showed that only positive UV boundary values of the cosmological and Newton constant, $\Lambda_{\rm cc}>0$ and $G>0$, are physically relevant (see comments below\,\eqref{anLambda} and\,\eqref{anG}). We also showed that $\Lambda_k$ ($=\Lambda_{\text{\tiny$L$}}$) and $G_k$ ($=G_{\text{\tiny$L$}}$)  do not change sign all along their flow, and obviously the same holds true for $\lambda$ and $g$. Therefore, the point $(\lambda_2,g_2)$ has to be excluded from the present analysis\footnote{For completeness, in the Appendix we will also consider the remaining (unphysical) cases.}, and we are left with the Gaussian fixed point $(\lambda_1,g_1)$ only. 
As anticipated, in contrast with the so-called asymptotic safety scenario \cite{Reuter:1996cp,Reuter:2001ag, Bonanno:2004sy}, our analysis does not show any sign of a non-trivial fixed point $\lambda_{\text{\tiny FP}}>0$\,, $g_{\text{\tiny FP}}>0$. 
Comments on the absence of such a fixed point are postponed to section \ref{Comparison}, where we will also explain what is at the origin of its appearance in \cite{Reuter:1996cp,Reuter:2001ag, Bonanno:2004sy}. 

\begin{figure}[t]
	\centering
	\includegraphics[width=0.6\linewidth]{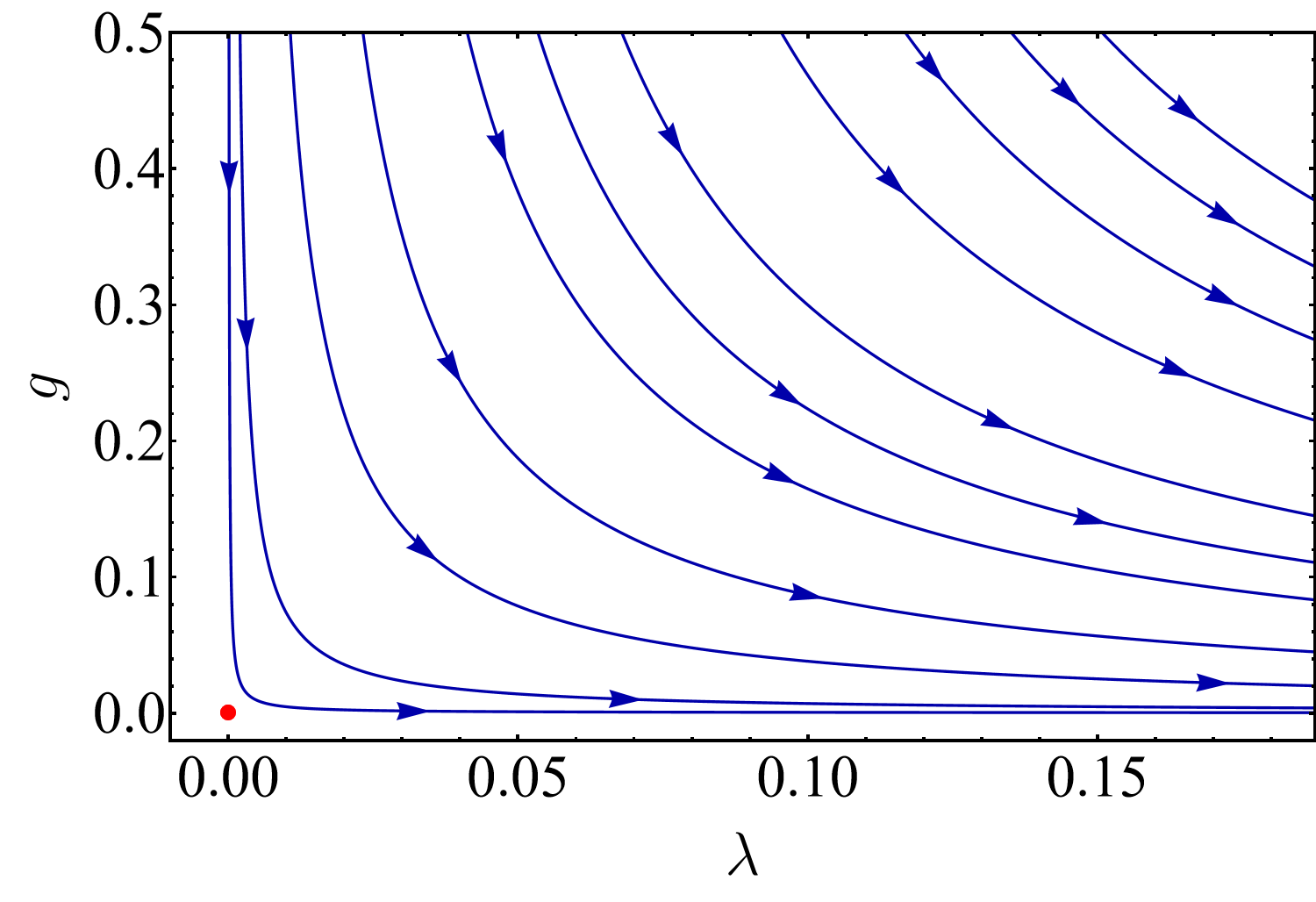}\hspace{15pt}
	\caption[]{\footnotesize RG flow from the numerical solution of\,\,\eqref{dimlessLambdaeq} and\,\eqref{dimlessGequation} in the physical quadrant $\left(\lambda>0,g>0\right)$.  The red dot is the Gaussian fixed point $(\lambda_1,g_1)=(0,0)$. The arrows point towards the IR.}
	\label{fluxnewlogic}
\end{figure}

From the analysis of the stability matrix 
\begin{equation}\label{M}
	M(\lambda,g)=\begin{pmatrix}
		\pdv{\beta_{\lambda}}{\lambda}& \pdv{\beta_{\lambda}}{g} \\
		\pdv{\beta_{g}}{\lambda} & \pdv{\beta_{g}}{g}
	\end{pmatrix}
\end{equation}
we find that $(\lambda_1,g_1)$ has a UV-repulsive eigendirection and a UV-attractive one, the axes $\lambda=0$ and $g=0$ respectively. 

As already said, the only physically relevant region is the quadrant $(\lambda>0\,, g>0)$, and we now move to a more complete study of the RG flow\,\eqref{dimlessLambdaeq}-\eqref{dimlessGequation} in this quadrant. Solving numerically these latter equations for different boundary conditions, we find the RG trajectories presented in Fig.\,\ref{fluxnewlogic}. The red dot is the Gaussian fixed point. The arrows point towards the IR and all the trajectories end at the minimal IR value of $\lambda$ allowed by\,\eqref{kIR}, namely $\lambda_{_{\rm IR}}=\Lambda_{4}/k_{_{\rm IR}}^2=3/16$. As already seen from the stability analysis, the $\lambda=0$ and $g=0$ axes are the corresponding UV-repulsive and UV-attractive eigendirections respectively.

In the previous section, we have seen that\,\eqref{Lambdaeq*} and\,\eqref{Gequation*} provide an excellent approximation to the RG equations\,\eqref{Lambdaeq} and\,\eqref{Gequation}. For this reason, it is worth to write the former equations also in terms of the dimensionless couplings $\lambda$ and $g$. We get
\begin{align}
	\dv{\lambda}{t}&=-2\lambda+\frac{6\lambda  g}{2\pi+3g}\label{dimlessLambdaeq*}\\
	\dv{g}{t}&=\,\,\,\,\,2g+\frac{6 g^2}{2\pi+3g}\,.\label{dimlessGequation*}
\end{align}
Eqs.\,\eqref{dimlessLambdaeq*} and\,\eqref{dimlessGequation*} can also be obtained directly from\,\eqref{dimlessLambdaeq} and\,\eqref{dimlessGequation} expanding the right hand side of both equations for $\lambda\ll1$ up to the first non-trivial order.
For boundary conditions $\lambda_0>0$ and $g_0>0$ at $t=0$, they admit the analytic solutions
\begin{align}
	\lambda&=\frac{\lambda_0}{2\left(1+\frac{\pi}{3g_0}\right)}\left[\,1+\sqrt{1+\frac{4\pi\,e^{-4t}}{3g_0}\left(1+\frac{\pi}{3g_0}\right)}\,\,\right]\label{dimlessLambda}\\
	g&=\frac{g_0\,e^{4t}}{2\left(1+\frac{\pi}{3g_0}\right)}\left[\,1+\sqrt{1+\frac{4\pi\,e^{-4t}}{3g_0}\left(1+\frac{\pi}{3g_0}\right)}\,\,\right]\,,\label{dimlessG}
\end{align}
that are nothing but  Eqs.\,\eqref{solutionLambda} and \eqref{solutionG} written for $\lambda$ and $g$. The solutions\,\eqref{dimlessLambda} and\,\eqref{dimlessG} very well approximate the numerical solution to the original equations\,\eqref{dimlessLambdaeq} and\,\eqref{dimlessGequation} (plotted in Fig.\,\ref{fluxnewlogic}) in the whole range  considered for $t$.

In view of the profound difference between our results and the asymptotic safety scenario \cite{Reuter:1996cp,Souma:1999at,Reuter:2001ag,Bonanno:2004sy}, that in the last decades has gained a certain popularity and has been considered for applications that range from black holes physics to inflation, we believe it worth to investigate on the origin of such a difference. This is the subject of the next section.

\section{Comparison with existing literature}
\label{Comparison}

The analysis that we performed in the previous sections did not show any sign of the UV-attractive fixed point of the asymptotic safety scenario found in \cite{Reuter:2001ag} (where the effective average action formalism was used) and later confirmed in \cite{Bonanno:2004sy} (resorting to the proper-time formalism). Pushing further our analysis, we now investigate on the reasons at the origin of the appearance of such a fixed point. 

To this end, we begin by considering our RG equation \eqref{RG3*} for the running action $S_{\text{\tiny$L$}}$, that for the reader's convenience we report below
\begin{align}
	L\pdv{}{L}S_{\text{\tiny$L$}}&=2\Lambda_{\text{\tiny$L$}}^2\,a^4+2\Lambda_{\text{\tiny$L$}}\left(L^2-8\right)a^2+\frac{L^4}{3}-\frac{34L^2}{3}+\frac{1859}{45}
	\,.
	\label{actionL}
\end{align}
As already explained, the relation between the numerical \vv running scale" $L$ and the physical running scale $k$ is given by $k=L/a^{\text{\tiny dS}}_{\text{\tiny$L$}}=L\sqrt{\Lambda_{\text{\tiny$L$}}/3}$ (see\,\eqref{relationRG} and\,\eqref{runningdS} and comments therein). For the purposes of the present analysis, we temporarily introduce the running scale $k$ in a different manner, namely through the relation 
\begin{equation}\label{incorrectk}
	k=\frac{L}{a}\,,
\end{equation}
i.e.\,\,we replace $a^{\text{\tiny dS}}_{\text{\tiny$L$}}$ in\,\eqref{relationRG} with the generic \vv off-shell" radius $a$ of the background metric $g^{(a)}_{\mu \nu}$. 

Obviously, this different identification of $k$ profoundly alters the $a$-dependence (powers of $a$) in the right hand side of\,\,\eqref{actionL}. In this respect, we recall that the running of\, $\frac{\,\Lambda_{\rm cc}}{8\pi G}$ \,and\, $\frac{1}{G}$ \,is determined once we insert the Einstein-Hilbert truncation\,\eqref{runningaction} in the left hand side of\,\eqref{actionL} and identify the coefficients of $a^4$ and $a^2$ of the first member with the corresponding ones of the second member (see comments above\,\eqref{Lambda} and\,\eqref{G}). Inserting\,\eqref{incorrectk} in \eqref{actionL} we then obtain the incorrect RG equation
\begin{align}\label{RGwrong1}
	k\pdv{}{k}S_k&=\left[\frac{k^4}{3}+2\Lambda_k \left(k^2+\Lambda_k\right)\right]\,a^4-\left(\frac{34k^2}{3}+16\Lambda_k\right)a^2+\frac{1859}{45}\,,
\end{align}
which gives rise to the (incorrect) equations ($\Lambda_{\text{\tiny$L$}}\equiv\Lambda_k$, $G_{\text{\tiny$L$}}\equiv G_k$)
\begin{align}
	\label{rhok}
	k\dv{}{k}\,\frac{\Lambda_k}{G_k}&=\frac{k^4+6\Lambda_k \left(k^2+\Lambda_k\right)}{\pi}\\
	\label{1suGk}
	k\dv{}{k}\,\frac{1}{G_k}&=\frac{17k^2+24\Lambda_k}{3\pi}\,,
\end{align}
that are easily translated in
\begin{align}
	k\dv{\Lambda_k}{k}&=\frac{G_k}{3\pi}\left(3k^4+k^2\Lambda_k-6\Lambda_k^2\right)\label{RGLambda}\\
	k\dv{G_k}{k}&=-\frac{G_k^2}{3\pi}\left(17k^2+24\Lambda_k\right)\,.
	\label{RGNewton}
\end{align}
 
These equations have to be compared with our correct RG equations\,\eqref{Lambdaeq} and\,\eqref{Gequation}. The two systems are profoundly different, and lead to significantly different results. Introducing the dimensionless running $\lambda$ and $g$ as in the previous section (see\,\eqref{dimlesscoup} and comments above), \eqref{RGLambda} and \eqref{RGNewton} are written as 
\begin{align}
	\dv{\lambda}{t}&=-2\lambda+\frac{g}{3\pi}\left(3+\lambda-6\lambda^2\right)\equiv\beta_{\lambda}\left(\lambda,g\right)\label{RGLambdaWrong}\\
	\dv{g}{t}&= 2g-\frac{\,\,\,g^2}{3\pi}\,\left(17+24\lambda\right)\equiv \beta_{g}\left(\lambda,g\right)\,.
	\label{RGNewtonWrong}
\end{align}
If we now search for the fixed points of this (incorrect) system, besides the Gaussian fixed point $(\lambda_1, g_1)=(0,0)$ we find the two other fixed points:
\begin{align}
	\left(\lambda_{\text{\tiny A}}, g_{\text{\tiny A}}\right)&=\left(\frac{\sqrt{154}-8}{30},\frac{2\pi}{23}\left(53-4\sqrt{154}\right)\right)\,\,\,\,\,\,\simeq\,\left(0.147,0.918\right)\label{FP1}\\ 
	\left(\lambda_{\text{\tiny B}}, g_{\text{\tiny B}}\right)&=\left(-\frac{8+\sqrt{154}}{30},\frac{2\pi}{23}\left(53+4\sqrt{154}\right)\right)\,\simeq\,\left(-0.680,28.039\right)\,.\label{FP2}
\end{align}
\begin{figure}[t]
	\centering
	\includegraphics[width=0.80\linewidth]{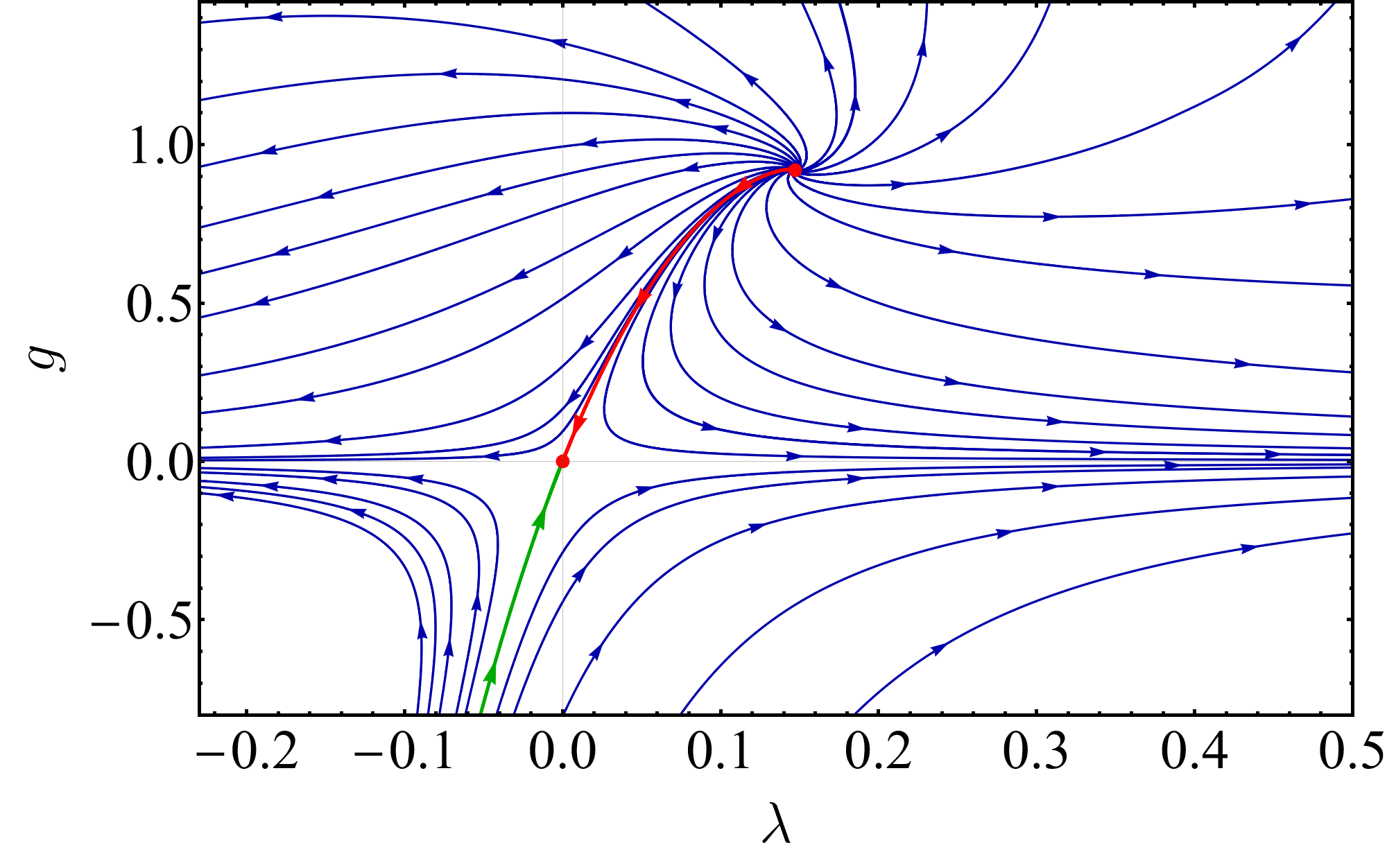}
	\caption[]{\footnotesize Flow from the numerical solution of the (incorrect) RG equations\,\eqref{RGLambdaWrong} and\,\eqref{RGNewtonWrong}. The red dots are the Gaussian fixed point $(\lambda_1,g_1)=(0,0)$ and the non-trivial UV-attractive fixed point $\left(\lambda_{\text{\tiny A}}, g_{\text{\tiny A}}\right)\simeq \left(0.147,0.918\right)$. In the half-plane $g>0$, the trajectories approach $\left(\lambda_{\text{\tiny A}}, g_{\text{\tiny A}}\right)$ in the UV with a spiralling behaviour. This is the typical flow of the asymptotic safety scenario to be compared with Fig.\,12 of \cite{Reuter:2001ag}. The red and green lines are the separatrix curves; the red one connects $\left(\lambda_{\text{\tiny A}}, g_{\text{\tiny A}}\right)$ to $\left(\lambda_1,g_1\right)$. Arrows point towards the IR.}
	\label{fluxwrong}
\end{figure}
 
Let us perform now the stability analysis for these fixed points\footnote{For all the fixed points, in the following we could write the exact (more involved) eigenvalues and eigendirections of the corresponding stability matrix (Eq.\,\eqref{M}), but this is not necessary as it does not provide any useful additional information. We report all the numbers approximated to the third digit.}. For the Gaussian fixed point, the stability matrix $M(\lambda,g)$ of Eq.\,\eqref{M} has a positive ($\theta_1=2$) and a negative ($\theta_2=-2$) eigenvalue, that correspond to a UV-repulsive ($g=4\pi\lambda$) and a UV-attractive ($g=0$) eigendirection. Similarly, $M\left(\lambda_{\text{\tiny B}},g_{\text{\tiny B}}\right)$ has a positive ($\theta_1\simeq28.453$) and a negative ($\theta_2\simeq-5.190$) eigenvalue, corresponding to a UV-repulsive ($g\simeq-65.741\,\lambda$) and a UV-attractive ($g\simeq627.551\,\lambda$) eigendirection.
Finally, for the fixed point $(\lambda_{\text{\tiny A}},g_{\text{\tiny A}})$ we find that the eigenvalues $\theta_1$ and $\theta_2$ of the stability matrix $M(\lambda_{\text{\tiny A}},g_{\text{\tiny A}})$ are $\theta_{1,\,2}\simeq-2.037\pm0.828 \,i$. This means that in the half plane\, $g>0$ \,the RG flow is UV-attracted towards this fixed point (negative real part) and has a spiralling behaviour in its vicinity (imaginary eigenvalues with opposite imaginary parts). Therefore, the point $(\lambda_{\text{\tiny A}},g_{\text{\tiny A}})$ is nothing but the fixed point of the asymptotic safety scenario \cite{Reuter:1996cp, Reuter:2001ag, Bonanno:2004sy}, and the analysis above shows that its appearance is due to the improper identification of $k$ through the relation\,\eqref{incorrectk}. In the following we will see that this identification is unfortunately what is implemented in the RG equations for $\Lambda_k$ and $G_k$ derived in \cite{Reuter:1996cp, Reuter:2001ag, Bonanno:2004sy}.

\begin{figure}[t]
	\centering
	\includegraphics[width=0.90\linewidth]{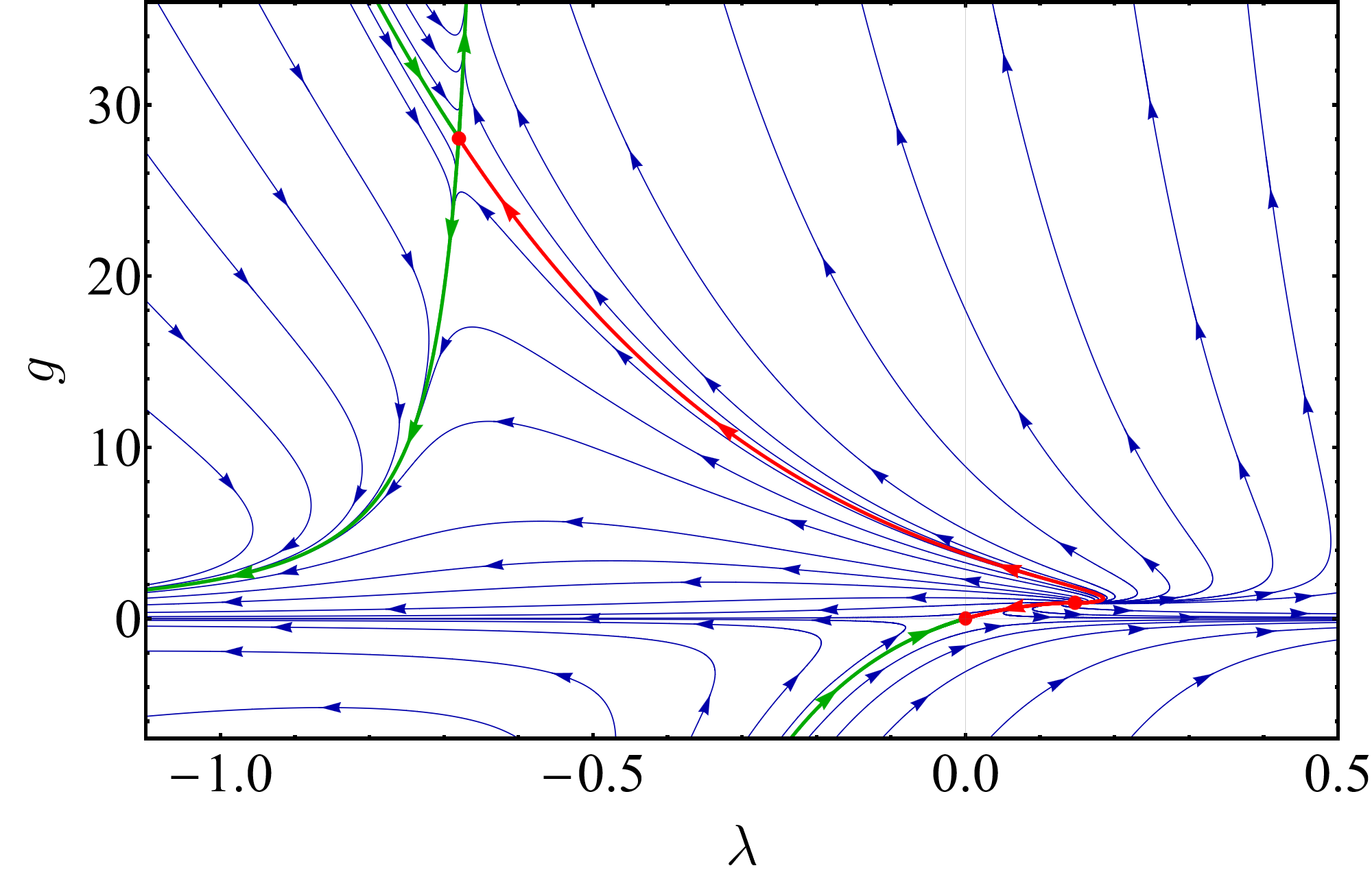}
	\caption[]{\footnotesize Same flow as in Fig.\,\ref{fluxwrong}, but with a wider range for the $\lambda$ and $g$ axes (to evidentiate the presence and the impact on the flow of the fixed point $(\lambda_{\text{\tiny B}},g_{\text{\tiny B}})$). The red dots are the fixed points $(\lambda_1,g_1)=(0,0)$, $(\lambda_{\text{\tiny A}},g_{\text{\tiny A}})=\left(0.147,0.918\right)$ and $(\lambda_{\text{\tiny B}},g_{\text{\tiny B}})=\left(-0.680,28.039\right)$. The red and green lines are the separatrix curves; the red ones connect the UV-attractive fixed point $(\lambda_{\text{\tiny A}},g_{\text{\tiny A}})$ to the other fixed points.}
	\label{fluxwrong2}
\end{figure}

Before doing that, we must stress another important result contained in Eqs.\,\eqref{RGLambdaWrong} and\,\eqref{RGNewtonWrong}. These equations, in fact, not only admit the non-trivial UV-attractive fixed point $(\lambda_{\text{\tiny A}},g_{\text{\tiny A}})$, but also provide an RG flow very similar to that of the asymptotic safety scenario. This can be immediately seen comparing our Fig.\,\ref{fluxwrong}, where we plot the RG trajectories obtained solving numerically the system\,\eqref{RGLambdaWrong}-\eqref{RGNewtonWrong} for different boundary conditions, with Fig.\,12 of \cite{Reuter:2001ag}. For the purposes of this comparison, in Fig.\,\ref{fluxwrong} we consider the same range of values for the $\lambda$ and $g$ axes of Fig.\,12 of \cite{Reuter:2001ag}. In Fig.\,\ref{fluxwrong2}, we show the same RG flow, but to better appreciate the impact on this flow of the fixed point $(\lambda_{\text{\tiny B}},g_{\text{\tiny B}})$ we consider a wider range\footnote{Clearly, as the fixed point $(\lambda_{\text{\tiny B}},g_{\text{\tiny B}})$ is located along the $g$ axis much above $(\lambda_{\text{\tiny A}},g_{\text{\tiny A}})$, in Fig.\,\ref{fluxwrong2} the portion of the RG flow shown in Fig.\,\ref{fluxwrong} appears to be flattened along this axis.} for the $\lambda$ and $g$ axes. In this respect, we note that a third fixed point as $(\lambda_{\text{\tiny B}},g_{\text{\tiny B}})$ does not appear in the plot presented in Fig.\,12 of \cite{Reuter:2001ag}. It should be stressed, however, that this plot was obtained with a specific choice of the regulator $R_k$ that appears in the RG equation for the effective average action (see\,\eqref{Reuter equation} below). Actually, it can be seen that different choices of $R_k$ can lead to the appearance of additional non-trivial fixed points other than the one of the asymptotic safety scenario. In particular, for some of these choices (actually for infinitely many choices) a non-trivial fixed point with the same characteristics of $(\lambda_{\text{\tiny B}},g_{\text{\tiny B}})$ is generated. Similar considerations apply also to the RG equations derived with the proper-time method, where again the results show a dependence on the cutoff function $f_k(s)$ used in this framework (see\,\eqref{PTeq} below).

To show now that within the effective average action formalism \cite{Reuter:1996cp, Reuter:2001ag}, as well as with the proper-time method \cite{Bonanno:2004sy}, the identification of the running scale $k$ in the derivation of the RG equations for $\Lambda_k$ and $G_k$ is realized through the improper relation\,\eqref{incorrectk}, we now reconsider the derivation of these equations with both these methods. 

{\it Effective average action} - Let us begin by considering the effective average action formalism. In this framework, the flow of $\lambda$ and $g$ is derived from the RG equation for the effective average action $\Gamma_k$. To see how this implementation of the RG transformation automatically incorporates the identification of $k$ through\,\eqref{incorrectk}, we now consider some relevant steps that lead to the RG equations of \cite{Reuter:1996cp,Reuter:2001ag}. The effective average action $\Gamma_k$ obeys the RG equation (as before, $t$ is the \vv RG time" $t=\log\frac{k}{k_0}$)
\begin{align}\label{Reuter equation}
	\partial_t\Gamma_k[g,\Bar g]=&\frac12 \Tr\left[\left(\kappa^{-2}\Gamma^{(2)}_k[g,\Bar g]+ R^{\rm grav}_k[\Bar g]\right)^{-1}\,\partial_t R_k^{\rm grav}[\Bar g]\right]\nonumber\\
	-& \Tr\left[\left(-\mathcal{M}[g,\Bar g]+R_k^{\rm gh}[\Bar g]\right)^{-1}\,\partial_t R_k^{\rm gh}[\Bar g]\right]\,,
\end{align}
where $\kappa\equiv\left(32\pi G\right)^{-\frac12}$, with $G$ bare Newton constant, $g_{\mu\nu}\equiv\Bar g_{\mu\nu}+\Bar h_{\mu\nu}$, with $\Bar g_{\mu\nu}$ a fixed gravitational background and $\Bar h_{\mu\nu}$ the classical field, and $\mathcal{M}[g,\Bar g]$ is the classical kinetic term of the ghosts
\begin{equation}
	\mathcal{M}[g,\Bar g]^\mu_{\,\,\,\,\nu}=\Bar g^{\mu\rho}\Bar g^{\sigma\lambda}\Bar D_\lambda\left(g_{\rho\nu}D_\sigma+g_{\sigma\nu}D_\rho\right)-\Bar g^{\rho\sigma}\Bar g^{\mu\lambda}\Bar D_\lambda g_{\sigma\nu}D_\rho\,,
\end{equation}
with $D_{\mu}$ and $\Bar D_{\mu}$ covariant derivatives that involve the Christoffel symbols for $g_{\mu\nu}$ and $\Bar g_{\mu\nu}$ respectively. The background metric $\Bar g_{\mu\nu}$ is eventually taken to be a sphere. $\Gamma^{(2)}_k[g,\Bar g]$ is the Hessian of $\Gamma_k[g,\Bar g]$ with respect to $g_{\mu\nu}$ at fixed $\Bar g_{\mu\nu}$. $R^{\rm grav}_k[\Bar g]$ and $R^{\rm gh}_k[\Bar g]$ are regulators that appear in the definition of $\Gamma_k$ (for the gravitational and ghost contribution respectively), both having the shape\footnote{We do not write here factors that are irrelevant to the present discussion. See \cite{Reuter:1996cp, Reuter:2001ag} for details.}
\begin{equation}\label{regulators}
	R_k[\Bar g]\sim \,k^2 R^{(0)}(-\square/k^2)\,\,,\quad\text{with}\quad \square\equiv\Bar g^{\mu\nu}\Bar D_\mu\Bar D_\nu\,,
\end{equation}
where $R^{(0)}(x)$ is a dimensionless function that interpolates between $R^{(0)}(0)=1$ and $\lim_{\,x\to\infty}R^{(0)}(x)=0$. In the effective average action method \cite{Wetterich:1989xg}, the cutoff functions $R_k$ implement the \vv Wilsonian" shell by shell elimination of modes, since they ensure that the (functional) traces in the right hand side of\,\,\eqref{Reuter equation} are effectively (i.e.\,\,in a smooth rather than a sharp sense) restricted to the eigenmodes of the Laplace-Beltrami operator $-\square$ whose corresponding eigenvalues $p^2$ lie \vv around" $k^2$:\, $p^2\sim k^2$. In the present case ($\Bar g_{\mu\nu}=g^{(a)}_{\mu\nu}$), this means that the running scale $k$ is identified with the eigenvalues of the Laplace-Beltrami operator  $-\square$ on the background sphere of radius $a$, i.e.\,\,through the relation\,\eqref{incorrectk}. 

The above considerations, the results found with the incorrect equations\,\eqref{RGLambdaWrong} and\,\eqref{RGNewtonWrong}, and the results of section\,\ref{fixedpoints} indicate that the UV-attractive fixed point of the asymptotic safety scenario does not exist and that its appearance is due to the improper identification of $k$ through\,\eqref{incorrectk}. Below we show that in the proper-time formalism \cite{Bonanno:2004sy} the same identification of $k$ is realized, and the same conclusions can be drawn also in this case.

{\it Proper-time method} - Let us move now to the proper time method. Indicating with $\Omega$ a typical fluctuation operator, the one-loop correction to the classical action in the proper-time representation is written as sum of contributions of the kind
\begin{equation}\label{ptomega}
	\Tr \log \Omega=-\Tr\int_0^{+\infty}\frac{\dd{s}}{s} e^{-s\,\Omega},
\end{equation}
where $s$, the so-called proper-time, is a parameter with dimension $({\rm mass})^{-2}$ and the UV divergence due to the lower bound of integration is regulated through the replacement $0\to1/\Lambda_{\rm cut}^2$. According to \cite{Bonanno:2004sy}, the Wilsonian RG strategy (shell by shell elimination of modes) is implemented in this framework considering the one-loop correction to the classical action, introducing an IR regulator $k$ that \vv\,suppresses contributions from large proper times $s\gtrsim k^{-2}$\,", i.e.\,\,making the replacement
\begin{equation}\label{hard cutoff}
	\int_{1/\Lambda_{\rm cut}^2}^{+\infty}\dd{s}\quad\longrightarrow\quad\int_{1/\Lambda_{\rm cut}^2}^{1/k^2}\dd{s}\,,
\end{equation}
taking the derivative with respect to $k$, and finally realizing the RG improvement of the one-loop result. All these steps lead to the RG equation for the gravitational action. A technical remark. In \cite{Bonanno:2004sy}, after presenting the hard cutoff implementation of\,\,\eqref{hard cutoff}, in the actual calculation the authors implement equivalent smooth cutoffs through the introduction in the proper-time integral of\,\,\eqref{ptomega} of functions $f_k(s)$ that smoothly interpolate between $f_k(s)\approx0$ \,for $s\gg k^{-2}$ \,and $f_k(s)\approx1$ for
$s\ll k^{-2}$ (similarly for the UV). Accordingly, they write the one-loop result for the gravitational action as 
\begin{equation}\label{oneloopPT}
	\widehat S_k[g,\Bar g]\equiv \widehat S\,[\Bar h; g]+\Gamma_1[g,\Bar g]_{\rm reg}\equiv\widehat S\,[\Bar h; g]-\frac12\Tr\int_0^{+\infty}\frac{\dd{s}}{s} f_k(s)\, \left[e^{-s\,\widehat S^{(2)}}-2\,e^{-s\,S_{\rm ghost}^{(2)}}\right]\,,
\end{equation}
where $g_{\mu\nu}$,\, $\Bar g_{\mu\nu}$ \,and\, $\Bar h_{\mu\nu}$ are as in\,\eqref{Reuter equation},\, $\widehat S\,[\Bar h; g]\equiv S[\Bar g+\Bar h]\,+\,S_{\rm gf}[\Bar h; \Bar g]$ (where $S$ is the classical action and $S_{\rm gf}$ the gauge-fixing term whose corresponding ghost action is $S_{\rm ghost}$), $\widehat S^{(2)}$ is the matrix of the second functional
derivatives of $\widehat S\,[\Bar h; g]$ with respect to $\Bar h_{\mu\nu}$, and likewise for $S^{(2)}_{\rm ghost}$.
Clearly, the hard cutoff of\,\,\eqref{hard cutoff} is implemented in\,\eqref{oneloopPT} choosing\, $f_k(s)=\theta\left(s-1/\Lambda^2\right)\theta\left(1/k^2-s\right)$. Finally, the RG equation for the running action $\widehat S_k[g,\Bar g]$ is obtained taking the derivative of both members of\,\,\eqref{oneloopPT} with respect to the \vv RG time" $t=\log\frac{k}{k_0}$ and eventually replacing in the right hand side {\footnotesize $\widehat S$} with {\footnotesize $\widehat S_k$}
\begin{equation}\label{PTeq}
	\partial_t\,\widehat S_k[g,\Bar g]=-\frac12\Tr\int_0^{+\infty}\frac{\dd{s}}{s} \,\partial_t f_k(s)\, \left[e^{-s\,\widehat S_k^{(2)}}-2\,e^{-s\,S_{\rm ghost}^{(2)}}\right]\,.
\end{equation}

To see that the RG equation\,\eqref{PTeq} implements the identification\,\,\eqref{incorrectk} for the running scale $k$, we now make the following observations. Taking the background metric $\Bar g_{\mu\nu}$ to be the metric $g^{(a)}_{\mu\nu}$ of a sphere of radius $a$, and considering for $\widehat S_k$ the Einstein-Hilbert truncation, we see that $\widehat S^{(2)}_k$ contains dimensionful Laplace-Beltrami operators $-\square$ for the sphere of radius $a$ (and different spins $0,1,2$) whose eigenvalues $\widehat \lambda_n$ go like $\widehat \lambda_n\sim\frac{n^2}{a^2}$. Moreover, in the right hand side of\,\,\eqref{PTeq} the term $\partial_t f_k(s)$ effectively selects the eigenmodes of $-\square$ whose corresponding eigenvalues lie in a narrow range (\vv infinitesimal shell") around $k^2$, i.e.\,\,$\widehat \lambda_n\sim k^2$. Therefore, as it is the case for the effective average action formalism, in the RG equation\,\eqref{PTeq} the running scale $k$ is identified through the relation\,\eqref{incorrectk}, and the same conclusions on the UV-attractive fixed point of the asymptotic safety scenario hold true.

In summary, we have shown that both the effective average action formalism and the proper-time RG implement the improper identification of the running scale $k$ through the relation\,\eqref{incorrectk}, and that the appearance of the non-trivial UV-attractive fixed point of the asymptotic safety scenario is due to this identification.

\section{Conclusions}
\label{conclusions}

In this work we considered the Einstein-Hilbert truncation for the running action in euclidean quantum gravity and, considering a spherical gravitational background, we derived the renormalization group equations for the running cosmological and Newton constant $\Lambda_k$ and $G_k$. We have shown that there are two crucial aspects to which attention must be paid in order to derive the RG equations. One concerns the measure in the path integral that defines the running action. This measure contains terms coming from the integration over the conjugate momenta of the original Hamiltonian formulation of the theory that are often neglected or mistreated. The other aspect concerns the identification of the physical running scale $k$. If the latter is not correctly introduced, the RG flow is substantially altered. 

We have shown that in the usual implementation of the RG transformation, that typically resorts to the effective average action and/or to the proper-time formalism, the running scale $k$ is improperly introduced. This results in altered RG equations for the cosmological and Newton constant, that lead to the generation of the UV-attractive fixed point of the asymptotic safety scenario. Moreover, we have shown that in the physically relevant quadrant $(\lambda>0,g>0)$ only the Gaussian fixed point exists, with a UV-attractive and a UV-repulsive eigendirection.

\section*{Acknowledgments}
We thank R. Percacci and D. Zappalà for useful discussions.	
The work of CB has been supported by the European Union – Next Generation EU through the research grant number P2022Z4P4B “SOPHYA - Sustainable Optimised PHYsics Algorithms: fundamental physics to build an advanced society” under the program PRIN 2022 PNRR of the Italian Ministero dell’Università e Ricerca (MUR). The work of VB, FC and AP is carried out within the INFN project  QGSKY.

\section*{Appendix}
\label{negative}

As said in the text, although only positive UV boundary values for the running cosmological and Newton constant are physically relevant, $\Lambda_{\rm cc}>0$ and $G>0$, for completeness in this Appendix we consider (and speculate on) the case of generic boundary values. Let us begin with $\Lambda_{\rm cc}>0$ and $G<0$. As for the case $\Lambda_{\rm cc}>0$ and $G>0$, the relation between the \vv running scale" $L$ and the physical running scale $k$ is given by\,\eqref{relationRG}, and the RG equations for $\Lambda_k$ and $G_k$ are\,\eqref{Lambdaeq} and\,\eqref{Gequation}. Considering the corresponding equations\,\eqref{dimlessLambdaeq} and\,\eqref{dimlessGequation} for the dimensionless couplings $\lambda$ and $g$, reported below for the reader's convenience
\begin{align}
	\dv{\lambda}{t}&=-2\lambda+\frac{2 g\lambda\left(3-2\lambda\right)}{2\pi+g\left(3-2\lambda\right)}\equiv\beta_{\lambda}(\lambda,g)\label{dimlessLambdaeqA}\\
	\dv{g}{t}&=\,\,\,\,\,2g+\frac{2g^2\left(3-8\lambda\right)}{2\pi+g\left(3-2\lambda\right)}\equiv\beta_{g}(\lambda,g)\label{dimlessGequationA}\,,
\end{align}
we find the non-trivial fixed point 
\begin{equation}\label{NTFPA}
	(\lambda_2,g_2)=\left(0,-\frac{\pi}{3}\right)
\end{equation}
in the $(\lambda>0,g<0)$ quadrant. In the main text this fixed point was already found (see Eq.\,\eqref{NTFP}), but it was discarded as we were interested only in the physical quadrant $(\lambda>0,g>0)$.
Performing the stability analysis, we find that the matrix $M(\lambda_2,g_2)$ (see Eq.\,\eqref{M}) has two negative degenerate eigenvalues ($\theta_{1,2}=-4$). To study the behaviour of the RG flow in the neighbourhood of $\left(\lambda_2,g_2\right)$ we have to linearize\,\eqref{dimlessLambdaeq} and\,\eqref{dimlessGequation} around this point. We find $\left(\lambda_2,g_2\right)$ to be a UV-attractive fixed point.

Let us consider now the case of negative UV boundary values for the cosmological constant, $\Lambda_{\rm cc}<0$. As stressed in section\,\ref{solutions}, the running cosmological constant $\Lambda_L$ cannot change sign along its flow, so that in this case it is $\Lambda_L<0$ in the whole range $4\leq L\leq N$. We introduce the running physical scale $k$ as\footnote{Note that $k$ varies in the range $[k_{_\text{\tiny IR}}, \mpl]$. From\,\eqref{relationneg}  we see that $k_{_{\rm IR}}=\sqrt{\frac{16|\Lambda_4|}{3}\,}$.} (see\,\eqref{relationRG})
\begin{equation}\label{relationneg}
	k=L\sqrt{-\frac{\Lambda_L}{3}}=L\sqrt{\frac{|\Lambda_L|}{3}}\,.
\end{equation}
\begin{figure}[t]
	\centering
	\includegraphics[width=0.9\linewidth]{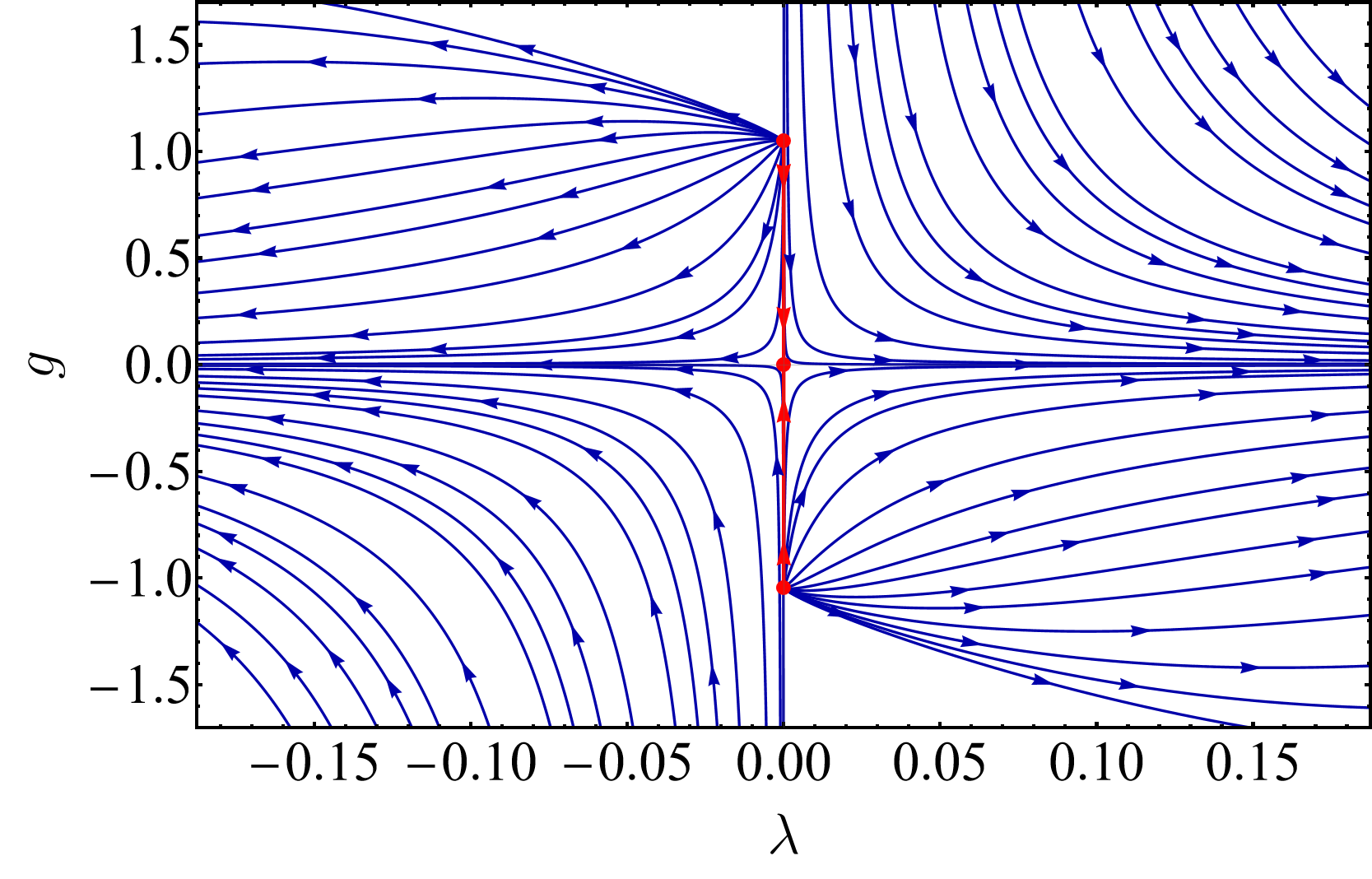}\hspace{15pt}
	\caption[]{\footnotesize RG flow in the whole plane $\left(\lambda,g\right)$ from the numerical solution of\,\,\eqref{dimlessLambdaeqA}-\eqref{dimlessGequationA} for $\Lambda_{\rm cc}>0$ and $G\gtrless 0$\,, and of\,\,\eqref{dimlessLambdaeq-}-\eqref{dimlessGequation-} for $\Lambda_{\rm cc}<0$ and  $G\gtrless 0$. The red dots are the three fixed points $(\lambda_1,g_1)=(0,0)$, $(\lambda_2,g_2)=\left(0,-\pi/3\right)$ and $(\lambda_3,g_3)=\left(0,\pi/3\right)$. The trajectories in the second quadrant $(\lambda<0,g>0)$ are UV-attracted by the fixed point $(\lambda_3,g_3)$, those in the fourth quadrant $(\lambda>0,g<0)$ by the fixed point $(\lambda_2,g_2)$.  The arrows point towards the IR, and the red lines connect the two non-trivial fixed points $(\lambda_2,g_2)$ and $(\lambda_3,g_3)$ to the Gaussian one. The trajectories in the second and fourth quadrant are symmetric with respect to $(0,0)$. The same holds true for the trajectories in the first and third quadrant. The flow in the first (physical) quadrant is the one plotted in Fig.\,\ref{fluxnewlogic}.}
	\label{fluxnewlogicunphys}
\end{figure}

\noindent
Inserting\,\eqref{relationneg} in\,\eqref{RGequations3_1} and\,\eqref{RGequations3_2} we finally get the RG equations ($\Lambda_k\equiv\Lambda_{\text{\tiny$L$}}$ and $G_k\equiv G_{\text{\tiny$L$}}$)
\begin{align}
	k\dv{\Lambda_k}{k}&=-\frac{3 G_k}{\pi}\frac{\Lambda_k\left(k^2+\frac23\Lambda_k\right)}{1-\frac{3G_k}{2\pi}\left(k^2+\frac23\Lambda_k\right)}\label{Lambdaeq-}\\
	k\dv{G_k}{k}&=-\frac{3 G_k^2}{\pi}\frac{k^2+\frac83\Lambda_k}{1-\frac{3G_k}{2\pi}\left(k^2+\frac23\Lambda_k\right)}\label{Gequation-}\,,
\end{align}
that introducing the \vv RG time"  $t$ and the dimensionless running cosmological and Newton constant $\lambda(t)$ and $g(t)$ as in section\,\ref{fixedpoints} (see\,\eqref{dimlesscoup}) can be written as
\begin{align}
	\dv{\lambda}{t}&=-2\lambda-\frac{2 g\lambda\left(3+2\lambda\right)}{2\pi-g\left(3+2\lambda\right)}\equiv\beta_{\lambda}(\lambda,g)\label{dimlessLambdaeq-}\\
	\dv{g}{t}&=\,\,\,\,\,2g-\frac{2g^2\left(3+8\lambda\right)}{2\pi-g\left(3+2\lambda\right)}\equiv\beta_{g}(\lambda,g)\label{dimlessGequation-}\,.
\end{align}
Beyond the Gaussian fixed point, we also find
\begin{align}
	\left(\lambda_3,g_3\right)&=\left(0,\frac{\pi}{3}\right)\,,\label{NTFP*}
\end{align}
that from the stability analysis turns out to be a UV attractive fixed point as $(\lambda_2,g_2)$.

Having now at our disposal the system\,\eqref{dimlessLambdaeqA}-\eqref{dimlessGequationA} for $\Lambda_{\rm cc}>0$ and $G\gtrless 0$\,, and the system\,\eqref{dimlessLambdaeq-}-\eqref{dimlessGequation-} for $\Lambda_{\rm cc}<0$ and  $G\gtrless 0$\,, we can now study the RG flow in the whole $(\lambda,g)$ plane. Solving numerically these equations, we find the RG trajectories presented in Fig.\,\ref{fluxnewlogicunphys}. The red dots are the three fixed points\,\eqref{GFP},\,\eqref{NTFPA} and\,\eqref{NTFP*}, namely the Gaussian fixed point $(\lambda_1,g_1)$ and the two non-trivial ones $(\lambda_2,g_2)$ and $(\lambda_3,g_3)$.
The red lines connect the two non-trivial fixed points to the Gaussian one. Arrows point towards the IR. All the trajectories (blue lines) in the half-plane $\lambda>0$ end at the minimal IR value of $\lambda$ allowed by\,\eqref{kIR}, namely $\lambda_{_{\rm IR}}=\Lambda_{4}/k_{_{\rm IR}}^2=3/16$. The trajectories in the half-plane $\lambda<0$ end at the minimal IR value of $\lambda$ allowed by\,\eqref{relationneg}, i.e.\,\,$\lambda_{_{\rm IR}}=-|\Lambda_{4}|/k_{_{\rm IR}}^2=-3/16$. The plot in Fig.\,\ref{fluxnewlogicunphys} shows what already found with the stability analysis: (i) the $\lambda=0$ and $g=0$ axes are the UV-repulsive and UV-attractive eigendirections for the Gaussian fixed point; (ii) the RG trajectories are UV-attracted by $(\lambda_2,g_2)$ in the quadrant $(\lambda>0,g<0)$ and by $(\lambda_3,g_3)$ in the quadrant $(\lambda<0,g>0)$.

\end{document}